\documentclass[11pt]{article}
\usepackage{makeidx}
\usepackage{multirow}
\usepackage{multicol}
\usepackage[dvipsnames,svgnames,table]{xcolor}
\usepackage{graphicx}
\usepackage{epstopdf}
\usepackage{ulem}
\usepackage{hyperref}
\usepackage{amsmath}
\usepackage{amssymb}
\author{Gianni}
\title{}
\usepackage[paperwidth=595pt,paperheight=841pt,top=70pt,right=70pt,bottom=70pt,left=70pt]{geometry}

\makeatletter
	{\par\setlength{\parindent}{#3}
	\setlength{\leftmargin}{#1}       \setlength{\rightmargin}{#2}%
	\advance\linewidth -\leftmargin       \advance\linewidth -\rightmargin%
	\advance\@totalleftmargin\leftmargin  \@setpar{{\@@par}}%
	\parshape 1\@totalleftmargin \linewidth\ignorespaces}{\par}%
\makeatother 

\begin{document}

\begin{center}
{\Large \textbf{Title} An empirical analysis of the dynamics of both individual galaxies and gravitational lensing in galaxy
clusters without dark matter}
\end{center}

\begin{center}
{\Large G. Pascoli }
\end{center}

\begin{center}
{\Large Email: \href{mailto:pascoli@u-picardie.fr}{pascoli@u-picardie.fr}}
\end{center}

\begin{center}
{\Large Facult\'{e} des sciences}
\end{center}

\begin{center}
{\Large D\'{e}partement de physique}
\end{center}

\begin{center}
{\Large Universit\'{e} de Picardie Jules Verne (UPJV)}
\end{center}

\begin{center}
{\Large 33 Rue Saint Leu,  Amiens, France}
\end{center}

{\raggedright
\textbf{Abstract}
}

The existence of the flat rotation curves of galaxies is still perplexing. The
dark matter paradigm was proposed long ago to solve this
conundrum; however, this proposal is still under debate. In this paper, we search
for universal relationships solely involving the baryonic density that incorporate both
galactic dynamics and gravitational lensing in galaxy clusters without
requiring dark matter. If this type of formula exists, we show that it is
possible that it can clearly indicate that dark matter is either
perfectly tailored to baryonic matter or, from a more radical point of
view, even perhaps useless. If the latter situation is true, then we must give
greater visibility to models such as modified inertia (MOND) or even modified gravity (MOG).

\vspace{10pt}

\section{Introduction}

Many astrophysical observations seem to suggest
the existence of a massive and non-baryonic component of
the universe called dark matter (DM) (Roszkowski et al., 2018).
In the domain of galaxies, the DM paradigm appears very flexible with two adaptable parameters
per galaxy, with $2n$ parameters for $n$ galaxies. In addition, various types of
theoretical DM profiles have been proposed (Bertone, 2010). This high flexibility represents
a great advantage compared to other models, such as modified Newtonian dynamics (MOND), which uses one
universal parameter (Milgrom, 1983). However, it is also its weakness because it makes the model
firmly non-predictive.

Another very perplexing aspect of DM is its intrinsic nature.
What is the composition of the DM that supposedly composes
approximately $85\%$ of the matter in the Universe? Fritz Zwicky first
used the term "dark matter" in the 1930s. By studying the so-called Coma galaxy
cluster, he concluded that the very large velocities of the galaxies he
measured implied that the cluster had much more mass than the visible matter
suggested (Zwicky, 1933). Since then---that is, for ninety years---the scientific
community has sought a DM particle. However, it seems very
difficult to identify particles with elusive properties, especially when the
range of masses and the cross sections are unknown quantities. Despite
this fact, theorists have envisioned a wide range of weakly interacting massive
particles (WIMPs), representing a leading class of candidates. These particles
are predicted in many SUSY models and can be observed in the Large Hadron
Collider (LHC). Unfortunately, to date, the LHC has found no signs of WIMP
DM (Boveia and Doglioni, 2018; Bertone and Tait, 2018). Other terrestrial searches have also been infructuous (Xenon Collaboration, 2018). On the opposite side in the astrophysical domain, it was
proposed that gamma-ray signals could act as DM tracers. However, 
observations with the Fermi Gamma-Ray Space Telescope in the Small Magellanic
Cloud have led to a stringent limit on the DM annihilation cross section
(Caputo et al, 2016). A promising path seems to be comparison of the
gamma-ray data coming from the Fermi Large Area Telescope (LAT) and the
gravitational lensing data from the Dark Energy Survey (DES) (Ammazzalorso et
al, 2020). These researchers found a significant cross correlation between the
positions of gravitational lenses, which are thought to trace DM, and
those of gamma-ray photons, which are potentially emitted when DM
self-destructs. Unfortunately, the statistical analysis of the data remains
 unconclusive because the cross correlation predominantly comes from blazars.
The possibility that these data arise in some small part from DM
remains an open question.

Apart from WIMPs as potential candidates, a few researchers
have attempted to explain the phenomenon of DM with the
introduction of a gas composed of very hypothetical particles of negative mass
that could pervade the Universe (Farnes, 2018). However, this proposal provides a worse solution than DM with WIMPs. One reason is that the
long-term stability of a disk of positive matter (a galaxy) placed in a halo of negative masses is questionable. A stronger second reason is that
with the creation of real particles with negative masses, the vacuum
could decay into a lower-energy state so that the vacuum would be unstable.
Other reasons to dismiss this bold idea of negative masses are exposed in (Socas-Navarro, 2019).

Ultimately, we can say that the direct proof of the existence of DM or very other exotic particles is
rather thin. Other realistic theories that strongly challenge the DM
paradigm are MOND (Milgrom, 1983, 2018, 2020; McGaugh, 2015, 2021) and modified gravity or MOG (Moffat, 2008). In contrast to the DM paradigm, whose predictive
power is very limited due to its extreme flexibility, both MOND and MOG
include only a very small number of parameters. We know that the predictive
power of a theory considerably increases when the number of parameters
strongly decreases. For instance, MOND with one universal parameter is
capable of accurately predicting the shape of the rotation curves in advance
of the very-low-density galaxies, while this type of prediction is not possible
in the framework of the DM paradigm (McGaugh, 2020).

Finally, we can add a remark on a semantic point of view suggested by
McGaugh (2015): ``The need for dark matter is often referred to
as the \textit{missing mass problem}. This terminology prejudices the answer.
More appropriately, the proper terminology should be the acceleration
Discrepancy’’. It is from this perspective that we conduct our
research.

More recently, we have proposed a phenomenological approach, the
$\kappa{}$ model, with a minimal number of ingredients. From this perspective, we aim to describe the phenomena starting solely from
observational data (Pascoli, 2022). Strictly speaking, the
$\kappa{}$ model is not a theory as can be MOND or MOG. The $\kappa{}$ model is based on an
empirical relationship, as simple as possible, built on the sole knowledge of
the observable quantity, i.e., the baryonic  density. At a later stage,
 a theory may be built. This methodology
appears natural for historical reasons. We find notable examples in the
history of science: the Kepler laws preceded Newtonian mechanics, and the
Rydberg formula preceded quantum mechanics. In contrast, whenever an
``additional’’ ingredient appearing from nowhere has been proposed to
explain an experiment or an observation (phlogiston, the aether, the Vulcan
planet, etc.), a few decades later, the scholarly community realized its
uselessness. This is also possibly the fate expected for DM.

In the present paper, we define the baryonic data, essentially the
surface density deduced from the measurement of the luminosity of galaxies, as
intrinsic parameters. On the other hand, the DM parameters are
defined as extraneous parameters. Is it possible to obtain an
agreement between the prediction of a phenomenological model based on the
observational data and galactic dynamics by using solely a set of
intrinsic parameters, all extraneous parameters, especially DM, being
excluded from the list? This proposal is supported by the $\kappa{}$ model.
The basis relationship links the mean density measured by the terrestrial
observer to an intrinsic parameter called $\kappa{}$ (Pascoli, 2022). This
intrinsic parameter acts as a simple factor for both the inertia term and the
active mass or the gravitational constant $G$ in the dynamics equation. Let us specify that $\kappa{}$ is
then a calculable quantity directly derived from the baryonic density and not
an ad hoc parameter, as in the case of DM.

The $\kappa{}$ model hypothesizes that for an observer at a given place (for
instance the Earth) the lengths and velocities, associated with any object, located
    at a very large distance, must be multiplied by a factor $\kappa{}$. This
factor is strictly positive-definite. At the Sun position in the Milky Way, i.e.,
in the solar system, $\kappa{}$ is by convention taken equal to unity. Then
the final result is that in the dynamics equation, the inertia
term is multiplied by $\kappa{}$, mimicking a MOND effect. On the other
hand, the gravitational constant $G$, or the active
mass $M,$ is multiplied by the inverse factor $\frac{1}{\kappa{}}$, (mimicking an MOG effect)
(Pascoli and Pernas, 2020; Pascoli, 2022). However, by
contrast to other proposed models or theories, in the $\kappa{}$ model, these
effects are assumed to be only apparent. In reality, the equations (Newton
or Einstein) and the physical constants, when locally considered, are left
unchanged, any observer $O$ being free to locally choose ${\kappa{}}_0=1$. For a
terrestrial observer, labelled $\odot$, looking at a distribution of matter with
a maximum ${\rho{}}_c$, for instance, a galaxy, the factor $\kappa{}$ is
ultimately linked to the baryonic mean density $\rho{}$ by the following
logarithmic functional:

\begin{equation}
\frac{{\kappa{}}_c}{\kappa{}}=1+Ln(\frac{{\rho{}}_c}{\rho{}})
\end{equation}

The mean densities $\rho{}$, ${\rho{}}_c$ are estimated by the observer $\odot$.
This relationship is also accompanied by

\begin{align}
 \rho_c > \rho_\odot \ \ \ \    {\kappa{}}_c=1+Ln(\frac{{\rho{}}_c}{{\rho{}}_\odot})   \nonumber \\
 \rho_\odot > \rho_c \ \ \ \    \frac{1}{{\kappa{}}_c}=1+Ln(\frac{{\rho{}}_\odot}{{\rho{}}_c})
\end{align}

We can choose by convention ${{\kappa{}}_\odot}=1$, which is the value taken by $\kappa{}$ at the Sun
position in the Milky Way. Note that relations (1) and (2) are
empirical laws and must be simply admitted in the $\kappa{}$ model and are not derived
from it. Thus, other different forms of this type of logarithmic relation can
still be suggested under the set of physical conditions, for instance,
   the range of mean densities or if the matter is either concentrated in compact
objects, i.e.,  stars, or composed of a gaseous bulk. At this stage, we cannot resolve this point. From a practical
of view, the formula is expressed as a function of the baryonic surface
density $\Sigma{}$ (observable and measured) and the thickness $\delta{}$ along the line of
sight (undetermined, but adjustable), i.e.,

\begin{equation}
\frac{{\kappa{}}_c}{\kappa{}}=1+Ln(\frac{{\Sigma{}}_c\ \delta{}}{\Sigma{}\  {\delta{}}_c})
\end{equation}

A supplementary
ambiguity is that the thickness can be a function of the
radial distance $r$ in the galaxy. In the present paper, all the fits of the rotation curves are made
 assuming  $\delta{}=constant$ and for the inclination  $i=constant$ along a galactic radius. The results
supplied are thus given at a first-order approximation level,
and the rotation curves, which are provided here, are mean curves. Then, a more
accurate fit would be obtained by taking into account the variations in $\delta{}$ in Eq. (3). Thus, the great advantage of the
$\kappa{}$ model is that it can be refined with an increased precision
for the rotation curves starting from a universal relationship. Unfortunately,
the observational resources are limited for both the thickness and
 the inclination, and we consider this refinement premature at the moment. On the other
hand, we do not exclude the suggestion that some amount of unseen baryonic
matter very likely exists, for instance, in the form of rogue planets,
neutron stars or stellar black holes. A  pervasive gas composed of sterile  neutrinos can  also be invoked (Boyarsky et al, 2019). These ingredients, for which the
percentage is unknown, are not taken into account in the present paper
but contribute to increasing the baryonic surface density.

Along this path, the case of various spiral galaxies is examined,
and we eventually apply the $\kappa{}$ model to the Bullet Cluster problem.

\section{Rotation curves of the galaxies}

We begin first with a galaxy composed of a unique discoidal structure and then
examine galaxies composed of multiple structures.

\subsection{NGC 1560}

  The rotation curve of the low-density galaxy NGC 1560 has already been analysed in the framework of the $\kappa$ model (Pascoli, 2022). However, this analysis was very succinct, starting from the Newtonian curve supplied by Famaey and McGaugh (Famaey and McGaugh, 2012). Here, we re-examine the question to produce a mean rotation curve without the fluctuations but in a more self-consistent way. The surface density is displayed in Fig. 1. This type of galaxy is idealized as a simple
disc. A fit of the surface
density can be made with the unique exponential curve

\begin{equation}
\Sigma{}={\Sigma{}}_{cd}\ exp(-\frac{r}{r_d})
\end{equation}

{\raggedright
with ${\Sigma{}}_{cd}=32\ M_\odot\ {pc}^{-2}$ and $r_d=3\ kpc$. The rotation
curve for a thin exponential disk is usually ex-
}

{\raggedright
pressed by modified Bessel functions (Binney and Tremaine, 1987):
}

\begin{equation}
v^2\left(r\right)=4\pi{}G{\Sigma{}}_{cd}r_d{(\frac{r}{2r_d})\
}^2\left[I_1(\frac{r}{2r_d})K_1(\frac{r}{2r_d})-I_0(\frac{r}{2r_d})K_0(\frac{r}{2r_d})\right]
\end{equation}

Following the procedure admitted in the $\kappa{}$ model (Pascoli, 2022), the
velocity, which is measured by spectroscopy, is then given by

\begin{equation}
v_{meas}\left(r\right)=p_c^{0.5}{\left[1+Ln(\frac{{\Sigma{}}_{c\
}\delta{}}{\Sigma{}\ {\delta{}}_c})\right]}^{0.5} v(r)
\end{equation}

where the factor $p_c=\frac{1}{{\kappa{}}_c}$. This coefficient is calculated
using (2) with the reference ${\Sigma{}}_{\odot{}}=70\ {M_\odot}\ {pc}^{-2}$ and
${{\delta{}}_\odot}=500\ pc$. To simplify, we chose
$\delta{}={\delta{}}_c={\delta{}}_\odot$. Thus, for NGC 1560, $p_c=1.78.$  We
emphasize that no extraneous parameters are used in the fitting process.

Fig. 2 shows that the $\kappa{}$ model produces a mean velocity curve very similar to that of MOND. This result strongly suggests that for NGC 1560, the baryonic distribution
alone is sufficient to accurately predict the observed velocity curve, removing the need for DM. The LSB galaxies are in the low-density regime,
and following the $\kappa$ model, the magnification effect of the velocities is then reinforced. Note that the same circumstances arise with MOND because the LSB galaxies are in the low-acceleration regime. Therefore, it is not surprising that both MOND and the
$\kappa{}$ model provide very similar galactic rotation curves.

\begin{center}
\includegraphics[height=260pt, width=210pt]{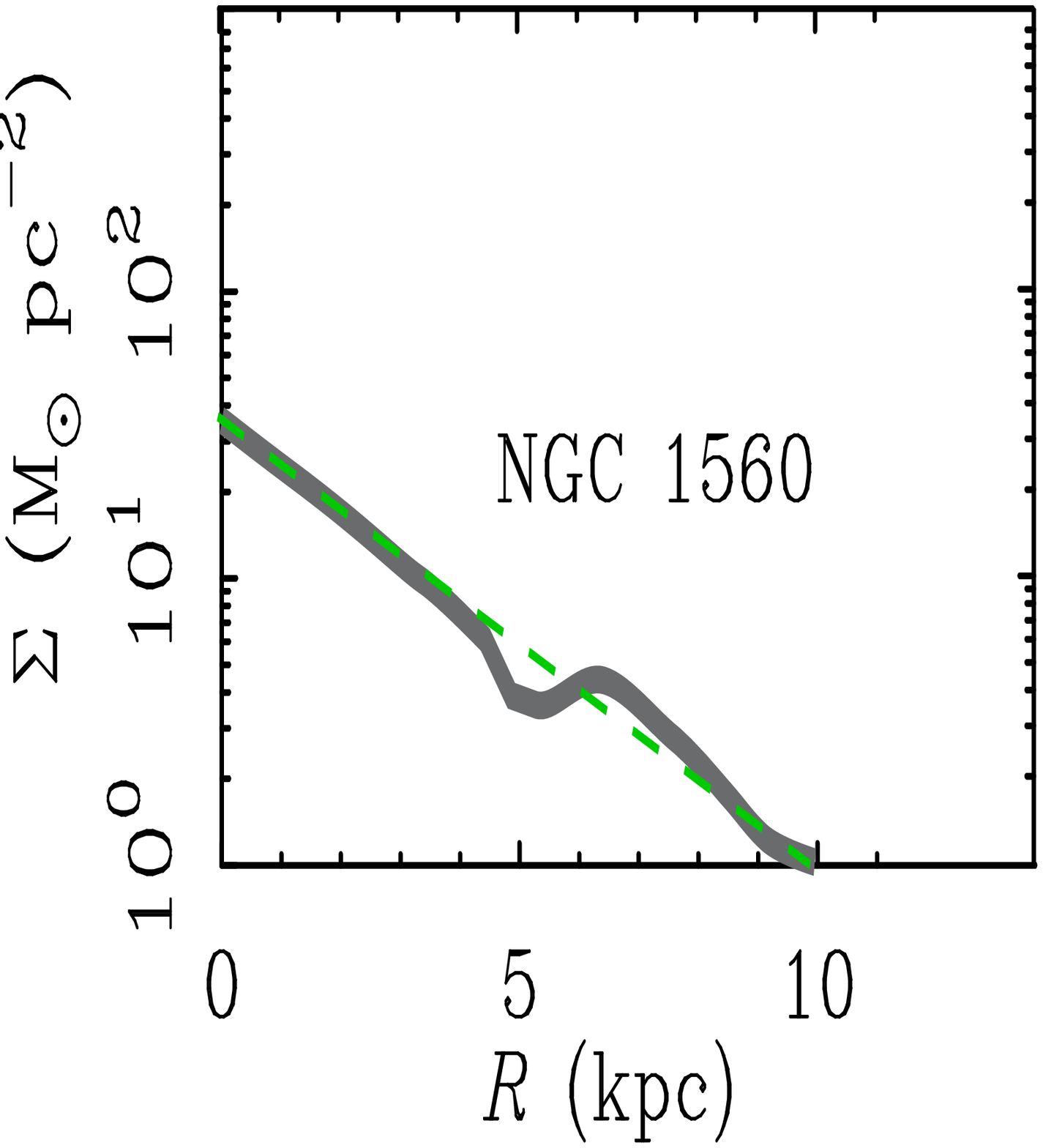}
\end{center}

Fig. 1 NGC 1560. The surface density profile is taken from Famaey and Mc Gaugh (2012). The green dashed line is the theoretical fit.

\begin{center}
\includegraphics[height=260pt, width=210pt]{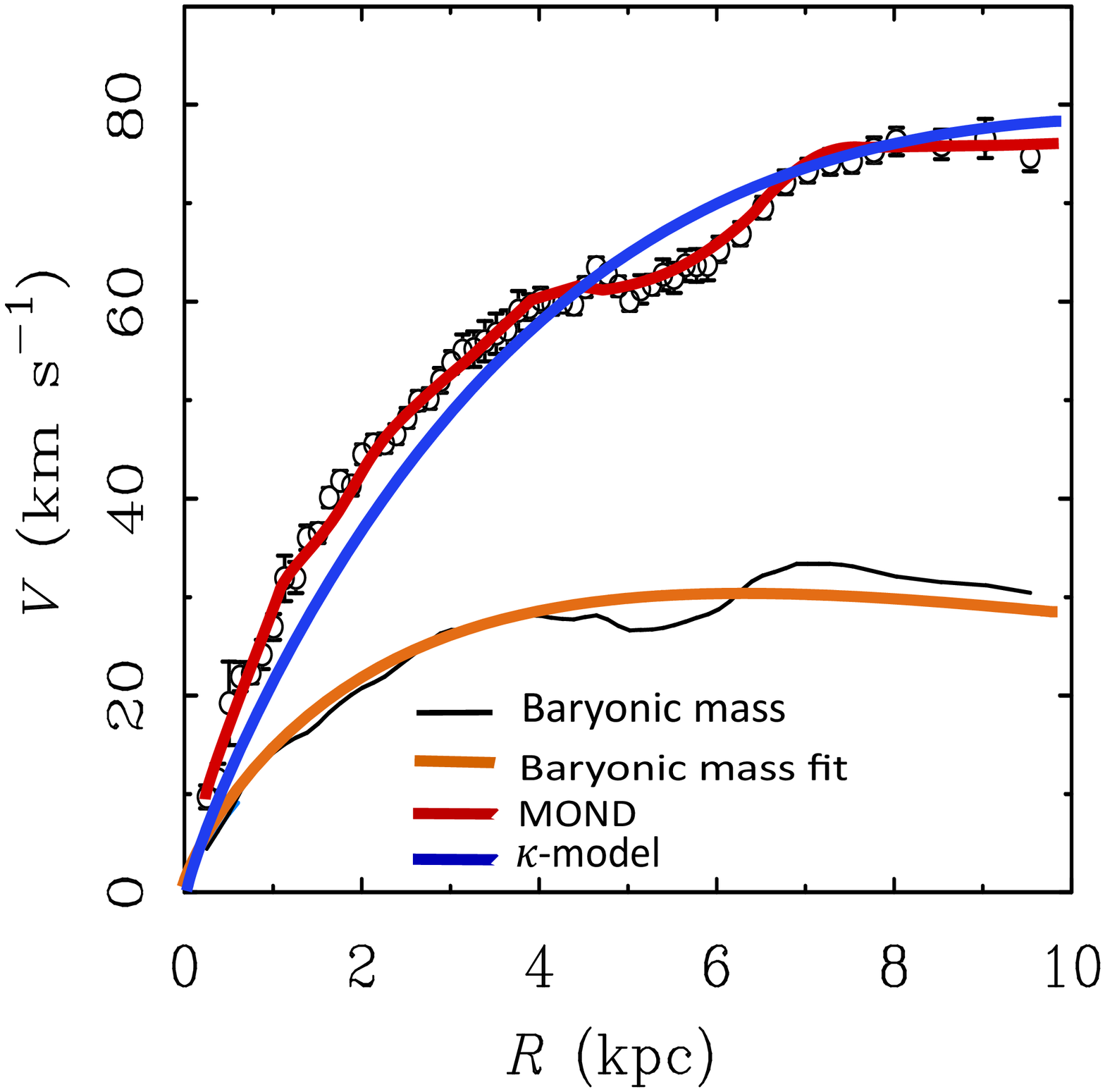}

Fig. 2 NGC 1560. Rotation velocity curves.
\end{center}

Note that analytic approximations starting from an exponential disc for the density are no longer valuable to reproduce the fine details (fluctuations) overlaid on the rotation curve. To move forward, we must necessarily take into account the inhomogeneities of the mass distribution. The method is then used to solve a Poisson equation to determine the gravitational potential that corresponds to the observed baryonic mass distribution. Nevertheless, to go beyond the first level of approximation, which directly leads to a mean rotation curve, we need to reach the thickness $\delta$ along the line of sight (Eq. 3). Knowing the surface density $\Sigma$ alone is not sufficient in the $\kappa$-model framework. Additionally, the inclination of the galaxy can also vary as a function of the radius $r$. These intermixed degeneracies are difficult to remove. Whereas the mean curve is easily reproduced by different theories (DM, MOG or MOND) or within the $\kappa$ model framework, a total interpretation of the rotation curve of a galaxy with its fluctuations remains ambiguous. This is true for a galaxy such as NGC 1560, which can be valuably fitted with one exponential disc but even more so for complex objects, as we now show.

\subsection{ Dwarf disc galaxies}

When the surface density cannot be fitted by a unique exponential profile, the analysis is more complicated. Karukes and Salucci (2017) studied a large sample of dwarf disc
galaxies located in the local volume. Those authors showed that a fit of
the surface density for these galaxies can be made with a sum of two
exponential (stars $s$ and gas $g$) profiles

\begin{equation}
\Sigma{}(r)=\frac{M_s}{2\pi{}
r_s^2}exp\left(-\frac{r}{r_s}\right)+\frac{M_g}{2\
\pi{}{(3r_s)}^2}exp\left(-\frac{r}{{3r}_s}\right)
\end{equation}

Normalizing this relationship immediately gives

\begin{equation}
\sigma{}(
x)={\sigma{}}_s\left[exp\left(-x\right)+\frac{\alpha{}}{9}exp\left(-\frac{x}{3}\right)\right]
\end{equation}

with two free intrinsic parameters ${\sigma{}}_s=\frac{{\Sigma{}}_s}{{\Sigma{}}_\odot}$ and
   $\alpha{}=\frac{M_g}{M_s}$ and $x=\frac{r}{r_s}.$ Fig. 7 of Kekures and Salucci (2017) is
reproduced in Fig. 3. The Newtonian curves for the stars (red line) and gas
(blue line) correspond to ${\sigma{}}_s=0.11$ and $\alpha{}=0.62.$ The $\kappa{}$-model profile, superimposed  in green, is in practice
confounded with the various DM profiles. The $\kappa{}$ model gives the
same mean rotation curve as the DM paradigm but without extraneous
parameters. One conclusion is clear: at least for the dwarf disc galaxies in
the local volume, the baryonic contribution is sufficient to explain the
rotation curves. Thus, the introduction of DM may be superfluous. We
can also take the point of view that the DM mimics an unknown emergent property of the baryonic matter itself at a cosmic macroscale level $\gtrsim 1 kpc$. Once again, these considerations give credence to simple models such
as MOND and the $\kappa{}$ model.

\begin{center}
\includegraphics[height=210pt, width=210pt]{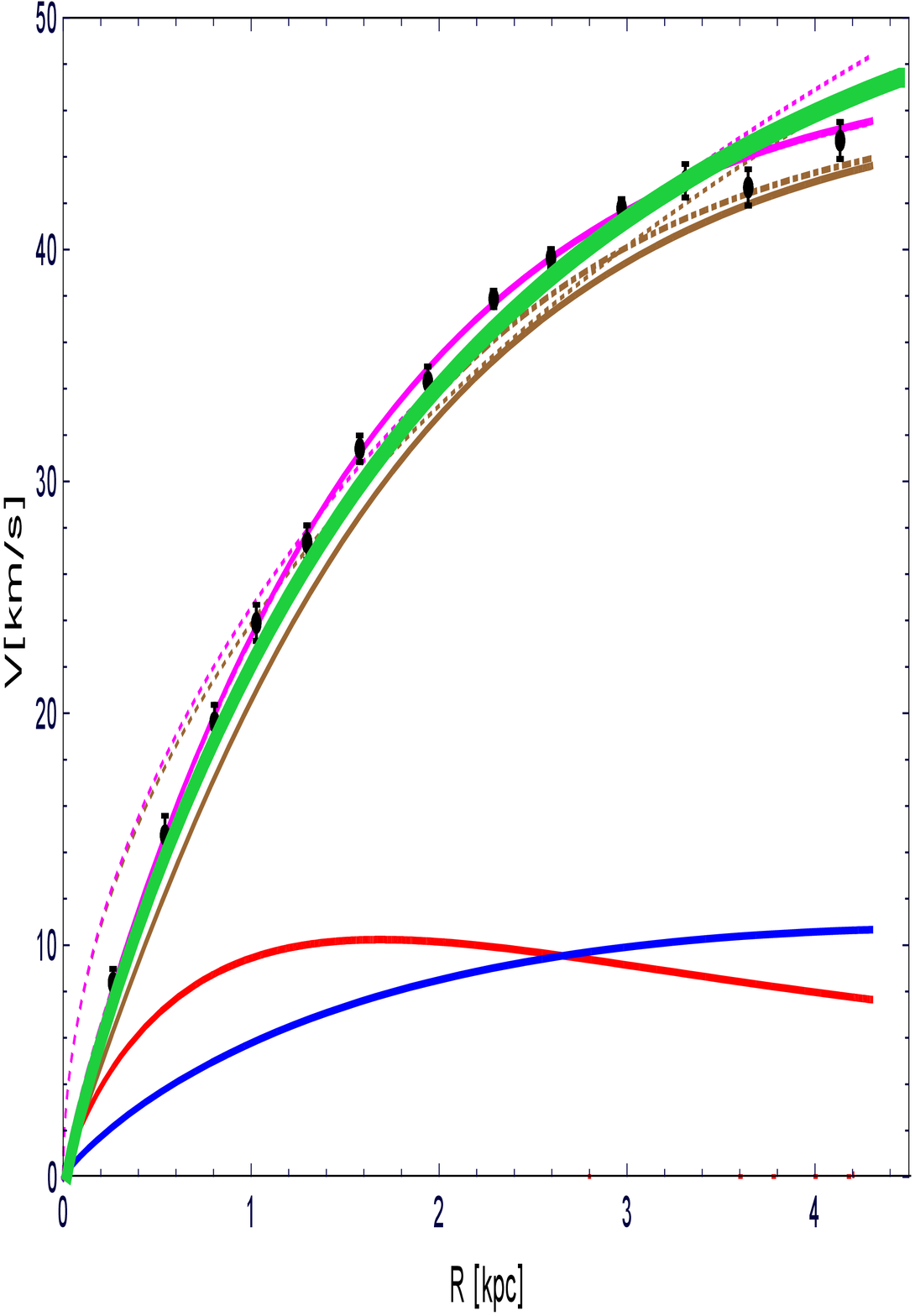}
\end{center}

Fig. 3 Synthetic rotation curve for dwarf galaxies (filled circles) and a set
of the theoretical curves (red line: stars; blue line: gas; brown line: halo; pink
line: the sum of all components) in the case of three DM profiles: the Burkert DM
profile (solid line), NFW profile (dotted line) and DC14 profile (dotted--dashed
line), from Kekures and Salucci (2017). The $\kappa{}$-model profile is superimposed
in green.

\subsection{NGC 6946}

 NGC 6946 is composed of multiple structures: a small bulge $b$, a stellar
disc $d1$ and a gaseous disc $d2$ (Fig. 3). The bulge can be modelled by a
 de Vaucouleurs-type profile (Binney and Tremaine, 1998)

\begin{equation}
{\Sigma{}}_{bulge}\left(r\right)={\Sigma{}}_{cb}exp\left[-8{\left(\frac{r}{r_b}\right)}^{\frac{1}{4}}\right]
\end{equation}

where ${\Sigma{}}_{cb}=6\ {10}^4\ $  $M_\odot\ {pc}^{-2}$, and $r_b=1\ kpc$. The volume mass density for the bulge is then calculated by the well-known formula

\begin{equation}
\rho{}\left(r\right)=\frac{1}{\pi{}}\int_r^{\infty{}}dx\
\frac{d{\Sigma{}}_b}{dx}\frac{1}{\sqrt{x^2-r^2}}
\end{equation}

{\raggedright
and the mass inside the radius $r$: $M\left(r\right)=4\pi{}\int_0^r dr'\ r'^2\
\rho{}\left(r'\right)$. This leads to a circular velocity of
}

\begin{equation}
v_b\left(r\right)=\sqrt{\frac{G\ M(r)}{r}}
\end{equation}

{\raggedright
where $G$ is the gravitational constant.
}

Each of the discoidal components is modelled with a Gaussian profile as

\begin{equation}
{\Sigma{}}_{disc}(r)={\Sigma{}}_{cd}exp\left[-{(\frac{r}{r_d})}^2\right]
\end{equation}

with $({\Sigma{}}_{cd1}=600\ M_\odot\ {pc}^{-2}$, $r_{d1}=3\ kpc)$ and
(${\Sigma{}}_{cd2}=10\ M_\odot\ {pc}^{-2}$, $r_{d2}=8\ kpc$)    (Fig. 4). The stellar and gaseous distributions of matter are then separately treated. We
affect the magnification factors ${p}_s=\frac{1}{{\kappa{}}_s}$ to
${\Sigma{}}_{bulge}\left(r\right)+{\Sigma{}}_{disc1}(r)$ and ${
p}_{d2}=\frac{1}{{\kappa{}}_{d2}}$ to ${\Sigma{}}_{disc2}(r)$. These coefficients
are calculated from the observational data of ${\Sigma{}}_{cb},$
${\Sigma{}}_{cd1}\ $ and ${\Sigma{}}_{cd2}.\ $ Numerically, we find ${
p}_s=0.129$ from ${\Sigma{}}_{cb}+$ ${\Sigma{}}_{cd1}$ and $p_g= 2.94$
from ${\Sigma{}}_{cd2}$. Ultimately, the rotation velocity must be multiplied by
the global factor
${\left\{1+Ln\left[\frac{{\Sigma{}}_{cb}/{\delta{}}_{cb}+{\Sigma{}}_{cd1}/{\delta{}}_{cd1}+{\Sigma{}}_{cd2}/{\delta{}}_{cd2}}{{\Sigma{}}_{bulge}\left(r\right)/{\delta{}}_{cb}+{\Sigma{}}_{disc1}\left(r\right)/{\delta{}}_{cd1}+{\Sigma{}}_{disc2}\left(r\right)/{\delta{}}_{cd2}}\right]\right\}}^{0.5}$.

\begin{center}
\includegraphics[height=260pt, width=210pt]{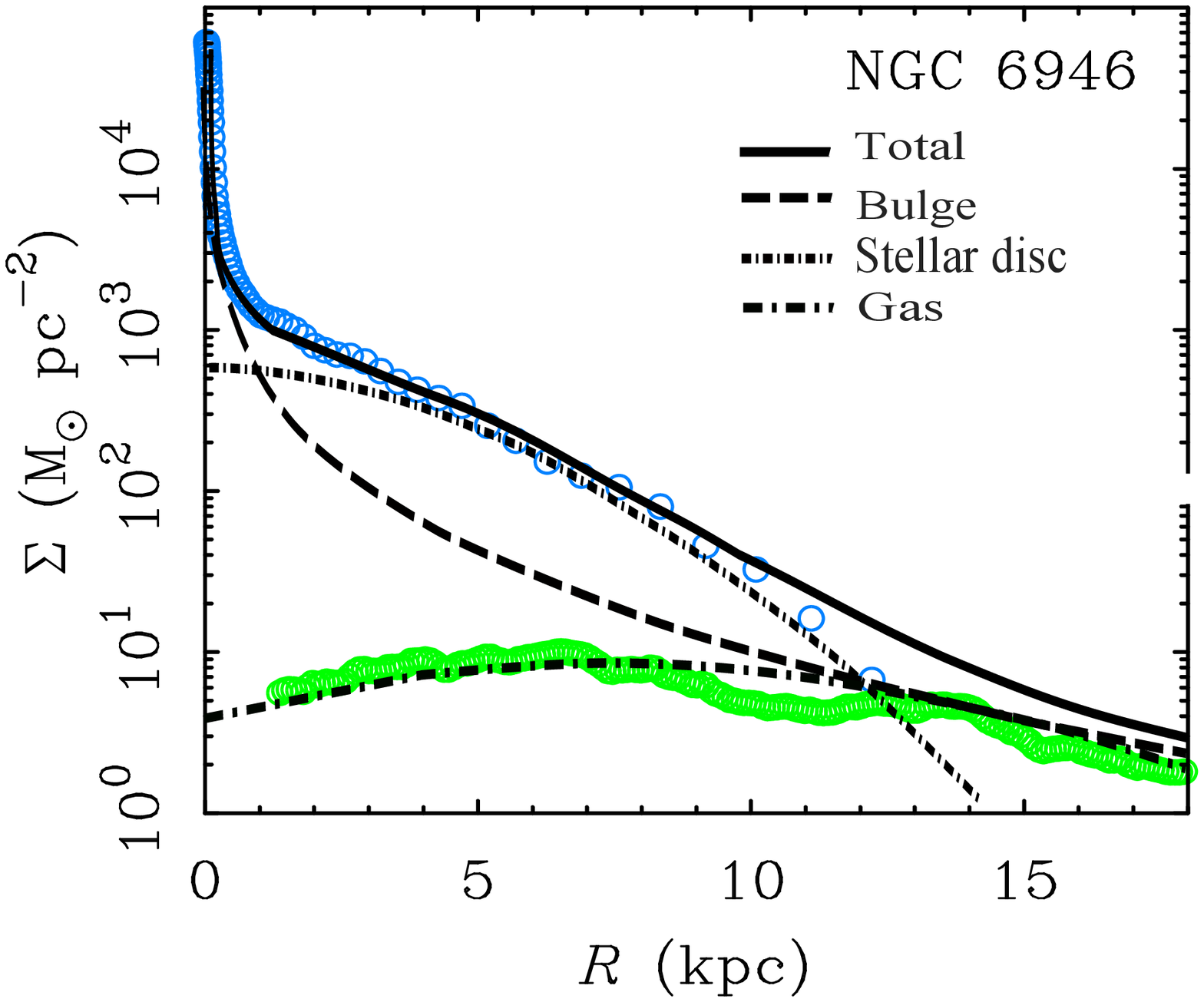}
\end{center}

Fig. 4 NGC 6946:    Observational surface density profile taken from Famaey and Mc Gaugh (2012) accompanied with the model.

\begin{center}
\includegraphics[height=260pt, width=210pt]{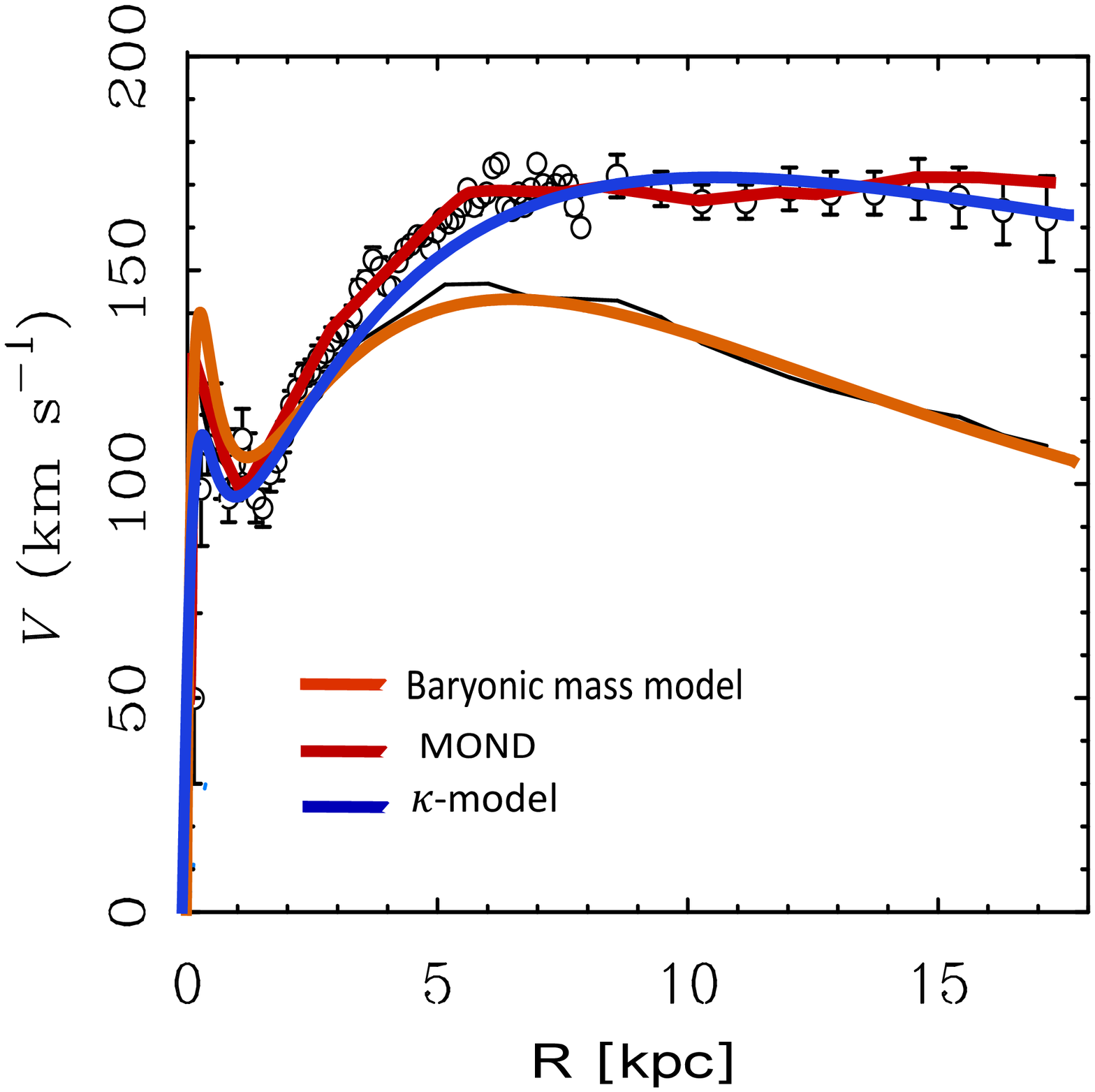}

Fig. 5 Rotation curves for $NGC 6946$.
\end{center}

\vspace{10pt}
Once again, we see  good agreement in the figure between MOND and the $\kappa$ model, together with the observational rotation curve.

\subsection{M33}

M33 is a low luminosity spiral galaxy in the Local Group. Its large angular
extent and well-determined distance make it ideal for a detailed study of the
radial distribution of matter (Corbelli, 2003). The surface density model
adopted here is composed of a stellar disc with an exponential fit
${\Sigma{}}_s(r)$ (${\Sigma{}}_{cs}=692\ M_\odot\ {pc}^{-2}$, $r_{ds}=1.1\ kpc$) and a
gaseous disc with a Gaussian fit ${\Sigma{}}_g(r)$
(${\Sigma{}}_{cg}=15\ M_\odot\ {pc}^{-2}$, $r_{dg}=7.6\ kpc$) (Fig. 6). The velocity is usually multiplied by the magnification factor
${\left\{1+Ln\left[\frac{{\Sigma{}}_s/{\delta{}}_s+{\Sigma{}}_g/{\delta{}}_g}{{\Sigma{}}_s\left(r\right)/{\delta{}}_s+{\Sigma{}}_g\left(r\right)/{\delta{}}_g}\right]\right\}}^{0.5}$.
A good fit, comparable to MOND (Fig. 7), is obtained for
${\delta{}}_g={\delta{}}_s={\delta{}}_{\odot{}}$ with the inclination of
54$^\circ{}$ chosen by Corbelli (2003). However, M33 appears clearly distorted, and
the inclination angle of M33 very likely varies in steps with the radial
distance $r$, as is conspicuous on the observational curve (Fig. 7).
Unfortunately, this parameter is difficult to estimate from observations.
Once again, the $\kappa{}$ model can give a more exact fit by adopting a
variable inclination from $60^\circ{}$ at 5 kpc to $48^\circ{}$ at 15 kpc.
Potentially, the $\kappa{}$ model can predict the variations in
both the inclination and the thickness along the line of sight. This statement is in turn a criterion of falsifiability of the $\kappa{}$ model.

\begin{center}
\includegraphics[height=220pt, width=220pt]{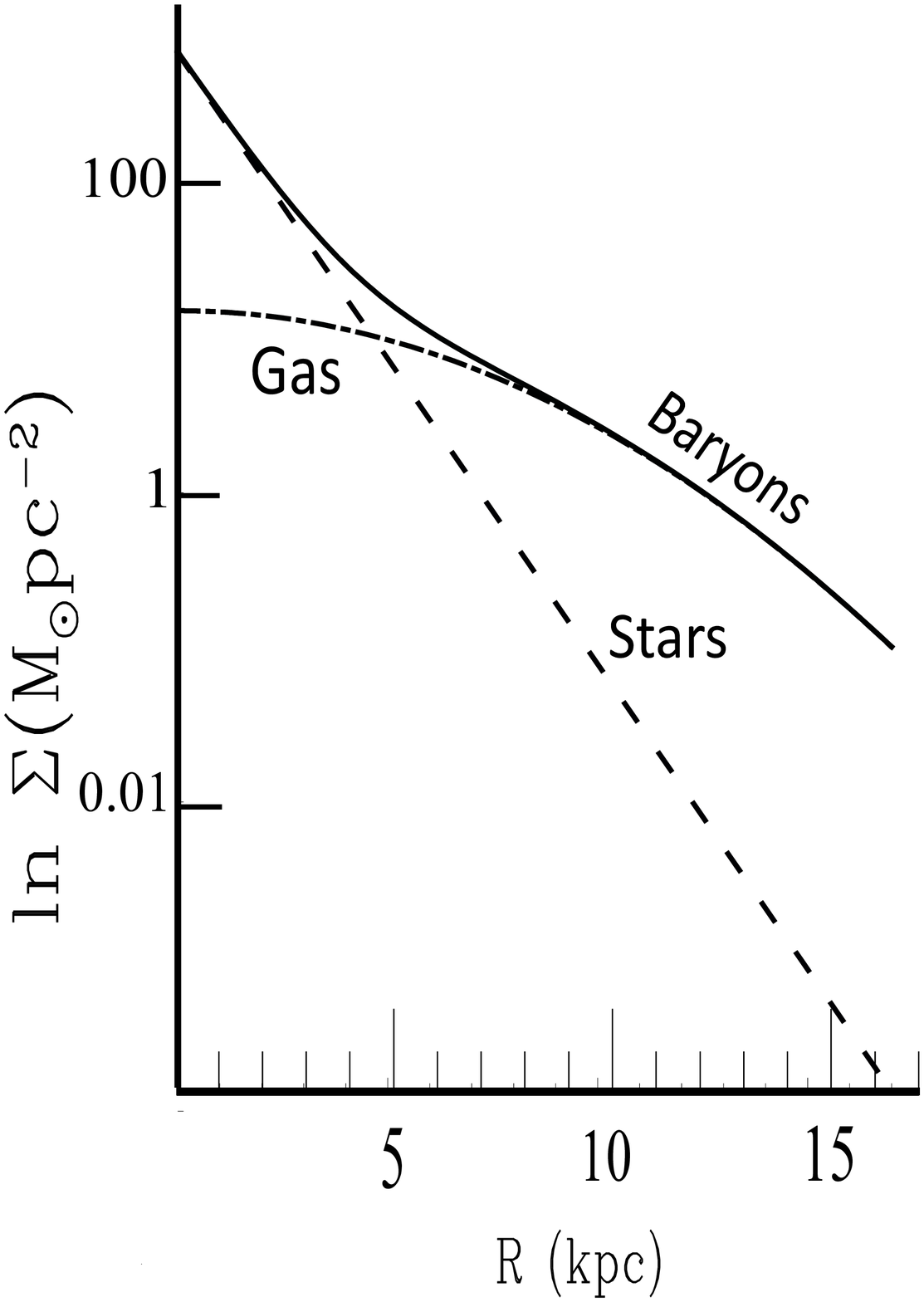}
\end{center}

Figure 6 The observational surface density profile (continuous line) is reproduced from Fig. 7 of Corbelli (2003) accompanied by the model described above.

\begin{center}
\includegraphics[height=260pt, width=310pt]{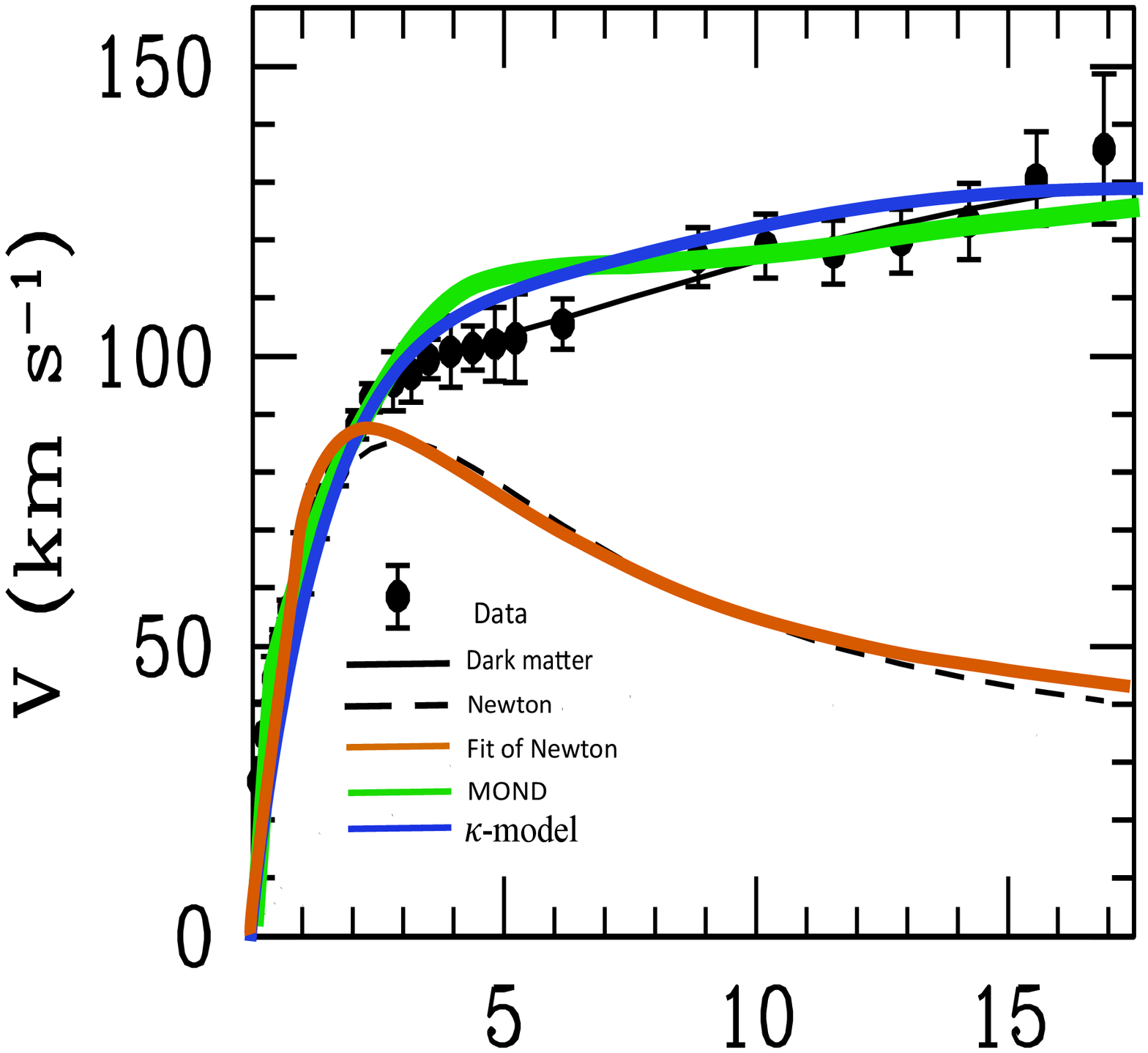}

Fig. 7 M33: Rotation curve decompositions

\end{center}

\subsection{Malin 1}

The surface density model adopted here is composed of a thick stellar disc
with an exponential profile ${\Sigma{}}_s(r)$ (${\Sigma{}}_{cs}=6\ {10}^3\ M_\odot\
{pc}^{-2}$, $r_{ds}=2\ kpc$) and a very extended thin gaseous disc with an exponential profile
${\Sigma{}}_g(r)$  (${\Sigma{}}_{cg}=4\ M_\odot\ {pc}^{-2}$, $r_{dg}=30\ kpc$) (Fig. 8).
With these values, the mass of Malin 1 is on the order of $1.7\ {10}^{11}\ M_\odot.\
$  Then, a good fit, similar to the predictions of DM or MOND, is obtained for
${\delta{}}_g={\delta{}}_s={\delta{}}_{\odot{}}$ but with a global change in
the inclination angle from $i= 38^\circ{}$ to $i=
34^\circ{}$. However, because the results vary with thickness in the
$\kappa{}$ model, a very similar fit is obtained with
$2{\delta{}}_g={\delta{}}_s={\delta{}}_{\odot{}}$ at constant $\Sigma{}$
without a change in the inclination $i = 38^\circ{}$. The rotation
curves are displayed in Fig. 9. The DM and the
$\kappa{}$-model profiles are very similar. The MOND profile appears slightly
less consistent in the outer regions than the two other profiles, but with a change in the
inclination angle in the outer parts of only $6^\circ{}$ from $i =
38^\circ{}$ to $i = 32^\circ{}$, which cannot be ruled out. According to
Lelli et al (2010), MOND is better than both the DM model and the
$\kappa{}$ model. Thus, the issue is not clearly resolved owing to the uncertainty
in the inclination in the outer region. In any case, both the thickness $\delta{}$
and the inclinations $i$ of the discs are variable quantities, which prevents
full exploitation of the observational data. Conversely,
the $\kappa{}$ model could supply predictive and falsifiable information on the mean thickness, but a fuller and more accurate set of observational points
would be needed.

\begin{center}
\includegraphics[height=220pt, width=210pt]{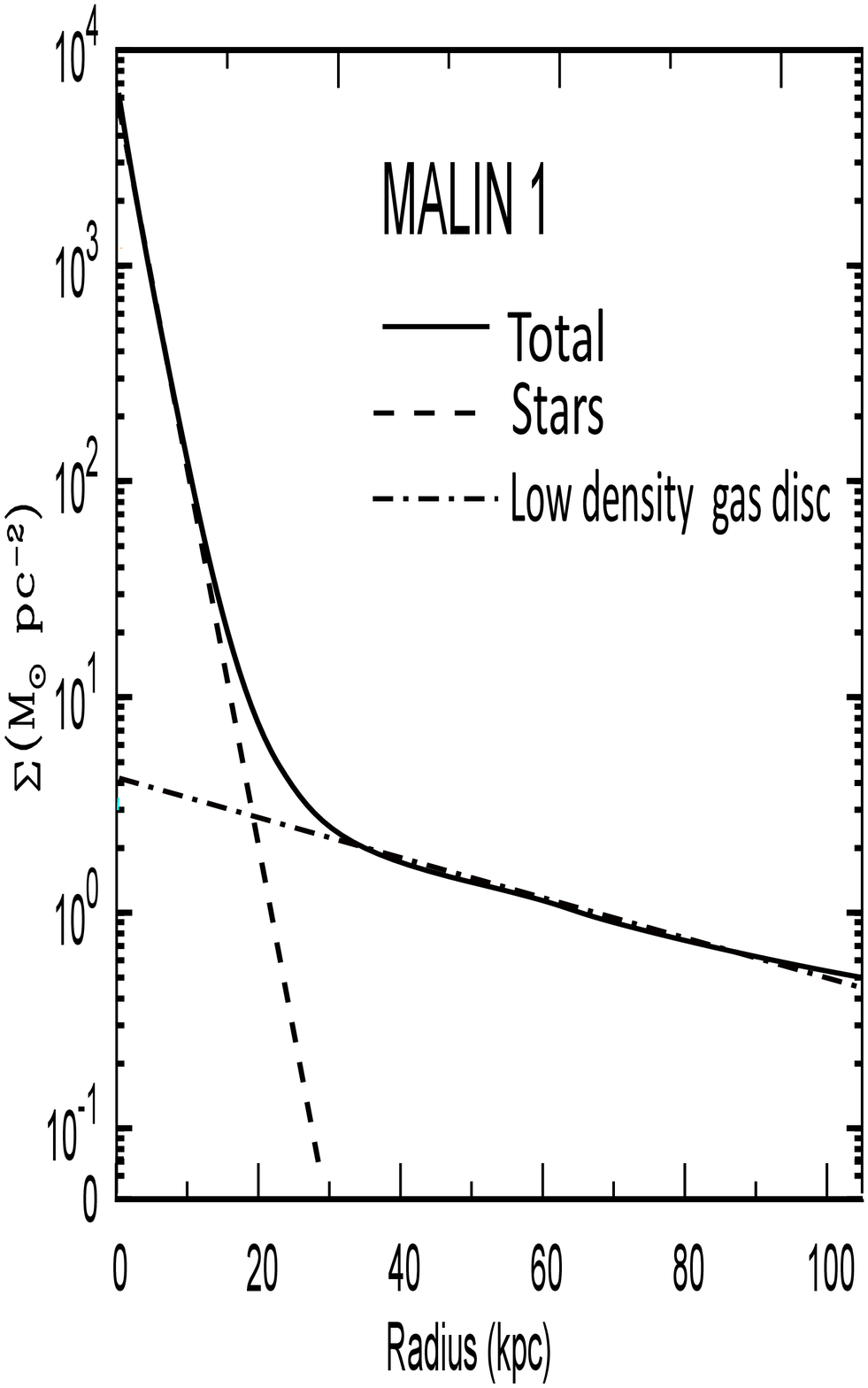}
\end{center}

Fig. 8 Observational surface density for Malin1 along a continuous line (from Lelli et al., 2010) and the model.

\begin{center}
\includegraphics[height=220pt, width=250pt]{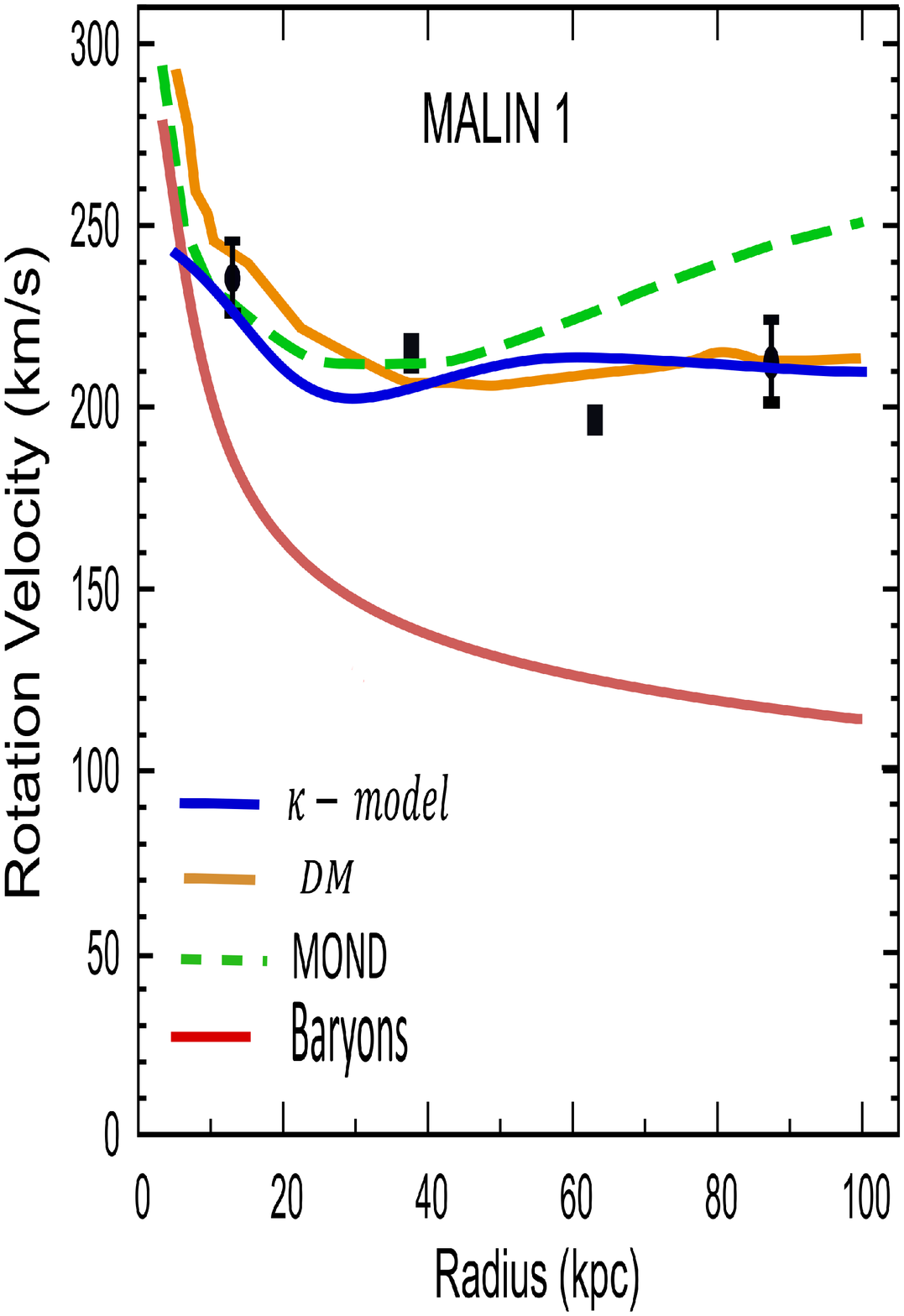}
\end{center}

Fig. 9 Malin1 Rotation curve decompositions. Dots with bar errors show the observed rotation curve (from Lelli et al, 2010). The $\kappa$-model curve appears in blue.

\subsection{NGC 7589}

The surface density profile is assumed to be decomposed into three
parts: a thick disc with an exponential distribution
${\Sigma{}}_s\left(r\right)$ (${\Sigma{}}_{cs}=3\ {10}^3\ M_\odot\ {pc}^{-2}$,
$r_{ds}=1\ kpc$), a thin HSB disc ${\Sigma{}}_{HSB}\left(r\right)$
(${\Sigma{}}_{cHSB}=400\ M_\odot\ {pc}^{-2}$, $r_{dHSB}=5.5\ kpc$) and a thin LSB
disc (${\Sigma{}}_{cLSB}=10\ M_\odot\ {pc}^{-2}$, $r_{dLSB}=15\ kpc$) (Fig. 10). A
good fit for the velocity curve is obtained with an inclination of
45$^\circ{}$. The latter value sensibly differs from that given by Lelli et al. (2010),
with $i=58^\circ{}$. The mean rotation curves are displayed in Fig. 11. Another similar solution is still to take
$3{\delta{}}_{LSB}=3{\delta{}}_{HSB}={\delta{}}_s={\delta{}}_{\odot{}}$ at
 constant $\Sigma{}$, while now maintaining the inclination
suggested by those authors. A mixture of intermediary solutions is clearly possible by exploiting both the inclination and\ the thickness. $NGC\ 7589$ provides a good example indicating that discrimination between different models can be challenging owing to the indeterminacy impacting these parameters when exploiting the
observational material.

\begin{center}
\includegraphics[height=220pt, width=210pt]{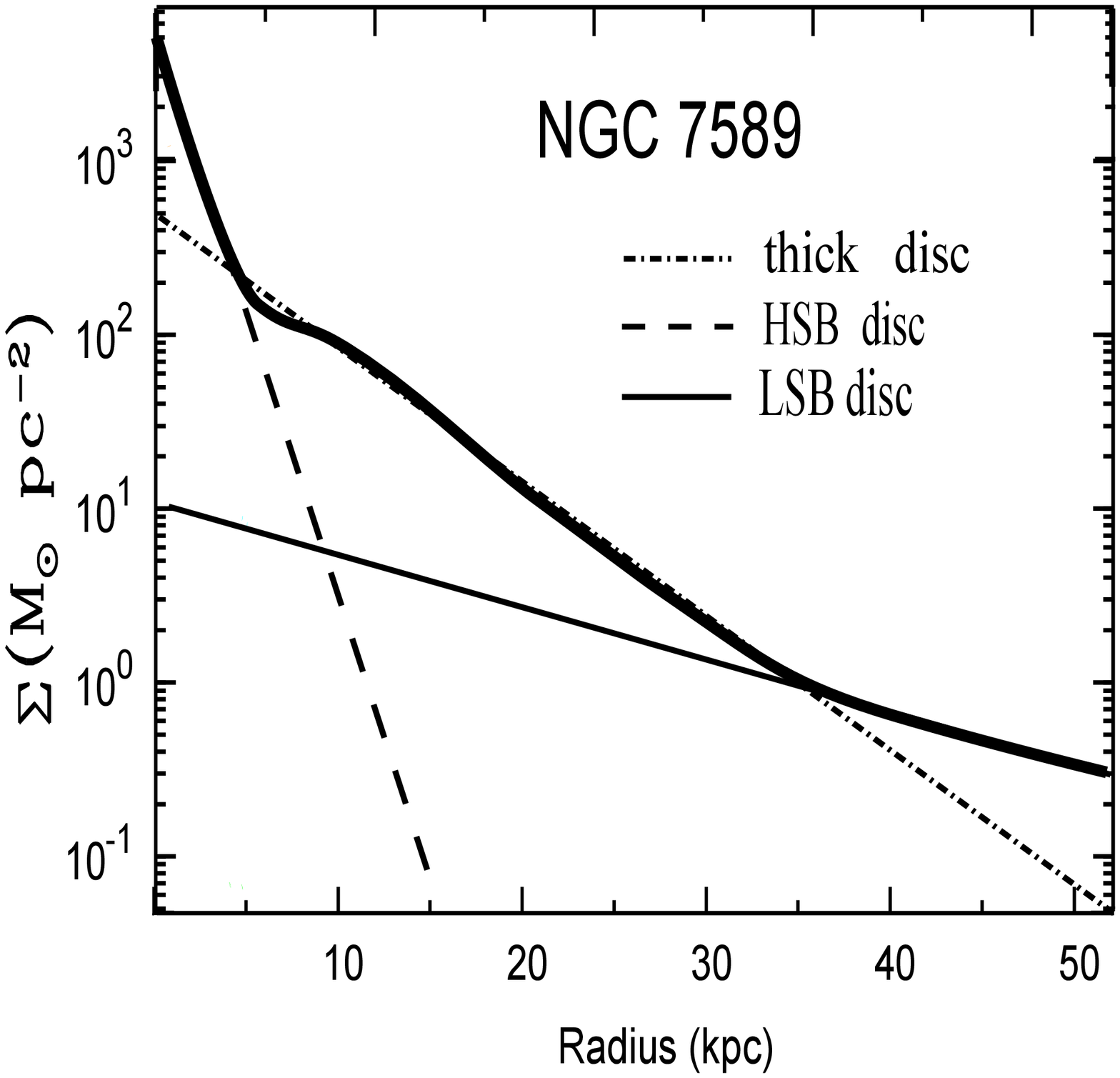}
\end{center}

Fig. 10 NGC 7589: Observational surface density along a thick continuous line (from Lelli et al., 2010) and the model.

\begin{center}
\includegraphics[height=220pt, width=210pt]{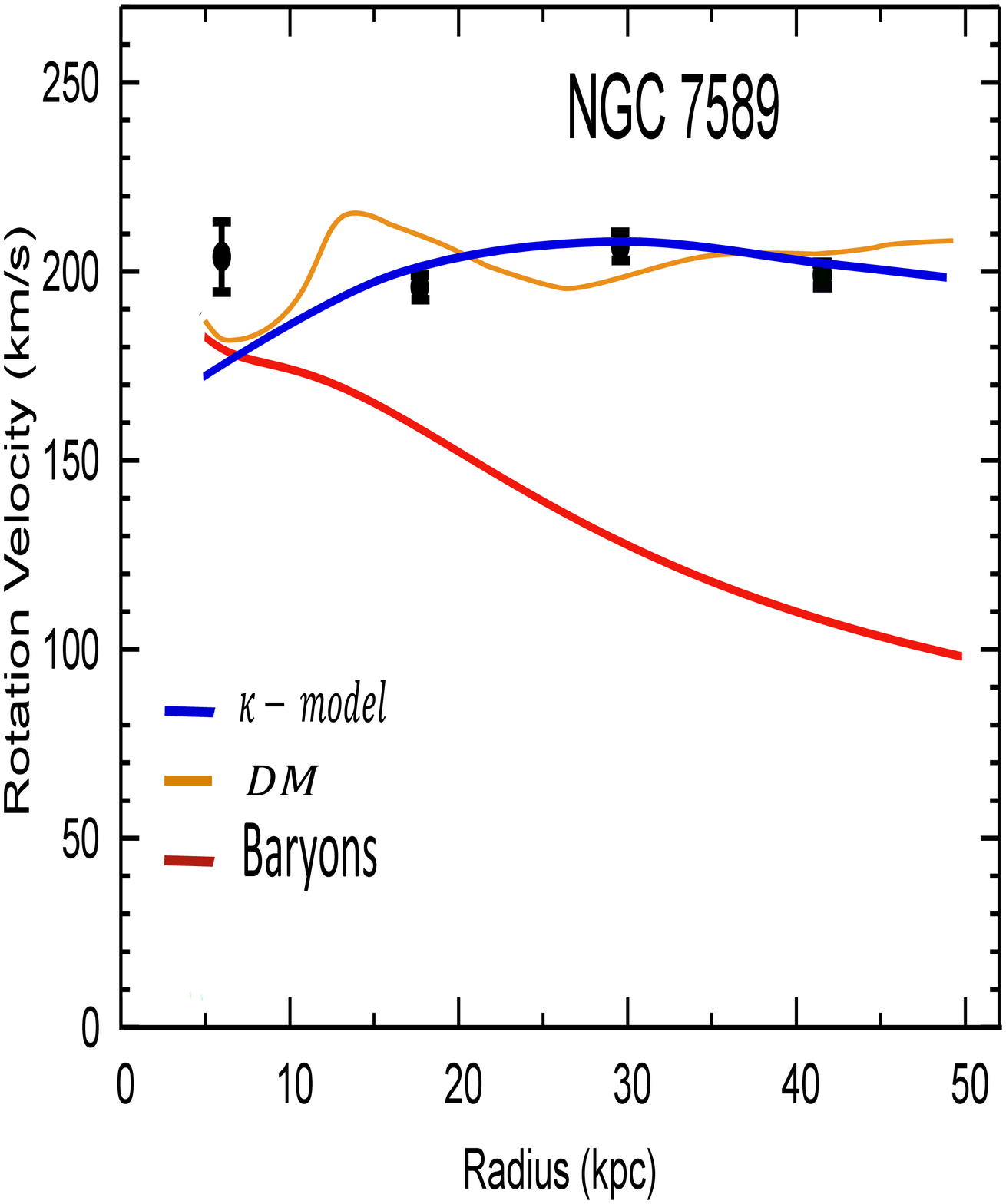}
\end{center}

Fig. 11 NGC 7589: Rotation curve decompositions. Dots with error bars show the observed rotation curve (from Lelli et al, 2010). The $\kappa$-model curve appears in blue.

\subsection{M31}

$M31$ (the Andromeda galaxy) is a near-twin of the Milky Way. Unlike the Milky Way
seen from the inside, $M31$ is seen from the outside, and both the morphology and
The velocity field are much easier to determine.

The surface density is assumed to be composed of three parts: a spherical bulge
with a de Vaucouleurs profile $\left({\Sigma{}}_{cb}=4\ {10}^3\ M_\odot\ {pc}^{-2}
,\ r_b=1.5\ kpc\right),$   a stellar disc ($s$)  and a gaseous disc  ($g$), with the last two represented by an exponential profile $({\Sigma{}}_{cds}=4\
{10}^2\ M_\odot\ {pc}^{-2},\ r_b=5\ kpc)$ and $\left({\Sigma{}}_{cdg}=5\ \ M_\odot\
{pc}^{-2}\,\\ r_b=30\ kpc\right).$ Note the surface density of
the gaseous disc is not uniform but strongly oscillates (Chemin et al, Fig. 16, 2009)
and its representation by an exponential profile is necessarily very crude; nevertheless, this simple procedure appears justified by obtaining the mean
rotation curve.

Integrating the distributions of matter gives a bulge mass of $2.5\
{10}^{10}\ M_\odot,$    a stellar disc mass of 6.2 ${10}^{10}\ M_\odot$ and a gas
disc mass of 1.7 ${10}^{10}\ M_\odot$, in good agreement with the observational
data (Chemin et al, 2009). The mean rotation curve produced by the
$\kappa{}$ model is displayed in Fig. 12 (caption inserted in the box) together
with a set of data given by Chemin et al. Once again, the $\kappa{}$-model and
DM profiles are similar. As already ascertained by the latter authors,
the peculiar shape of the rotation curve is not readily reproduced by mass
distribution models. The central velocity dip cannot be modelled by any
fits. Chemin et al. have also shown that between $r = 6\ kpc$ and $r =
27\ kpc$, a mean inclination of $(74.3\ \pm{}\ 1.1^\circ{})$ is derived for the
gas disc. A prominent gas warp is detected inside $r\ =\ 6\ kpc$, where the disc
appears less inclined, while another external warp detected beyond $r\ = 27\
kpc$ makes the disc more or less inclined. Two ring-like
structures are also observed around $r = 2.5\ kpc$ and $r = 4.7-5.7\ kpc$. A
wealth of other details can also be identified. It is very difficult to
consider all these peculiarities. The same conclusion applies to the
$\kappa{}$ model, given that the method we have used automatically leads to a
mean rotation curve. Chemin et al. also derived from their results that the
dynamical enclosed mass extrapolated at $r\ =159 \ kpc$ is $M_{Dyn} \sim{}
\ {10}^{12}\ M_\odot$, leading to a dark-to-baryonic mass ratio of $\sim{}10$. For
this extrapolated distance, the $\kappa{}$ model predicts a very similar
magnification factor for the mass, $\sim{}9$, but without the halo of
DM.

\begin{center}
\includegraphics[height=260pt, width=320pt]{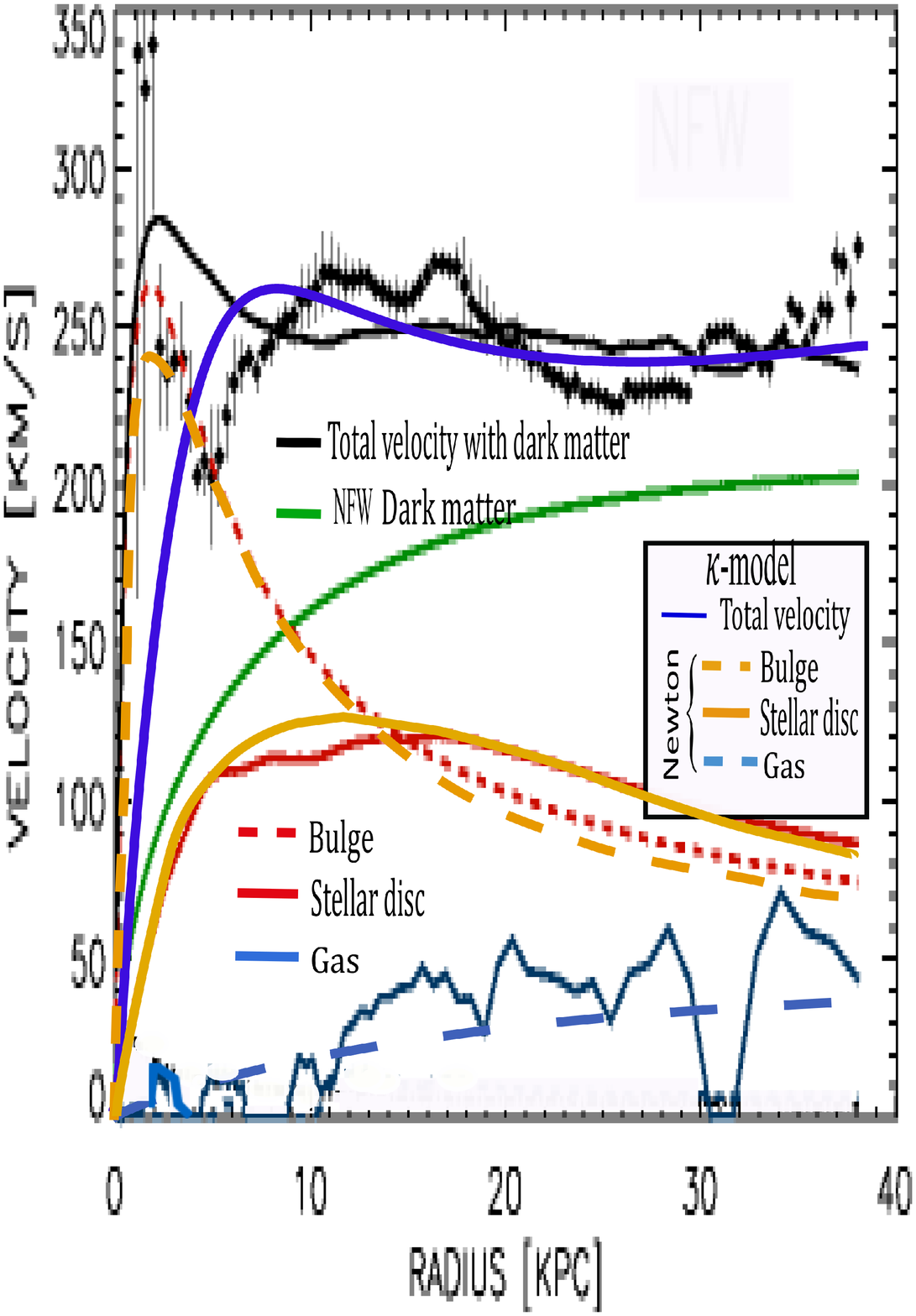}
\end{center}

\begin{center}
Fig. 12 $M31$: Rotation curve decompositions
\end{center}

We can conclude that starting from mean surface density profiles, the $\kappa{}$ model produces adequate mean velocity curves of galaxies, but
with the fluctuations smoothed. These fluctuations overlying the
observational velocity curves are not accurately taken into account by the
$\kappa{}$ model at the first level of approximation, nor by the DM
paradigm at the same level of approximation. These fluctuations are very likely
linked to heterogeneities in the repartition of masses and to variations as functions of the radius $r$ of both the inclination and the thickness along the line of sight. In any case, these
variations are very difficult to estimate, and some ambiguity is inevitable in the interpretation of the results, regardless of the model or
theory: DM, MOND, MOG, $\kappa{}$ model, \ldots{}. If the variation in
both the inclination and the thickness along the line of sight were
explicitly considered the $\kappa{}$-model could reproduce the fluctuations
seen on the rotation curves, but at the cost of degeneracies. Notably, this is also true for any other model.

\section{Bullet Cluster}

We now apply the $\kappa{}$ model to the Bullet Cluster by using the procedure proposed for the individual galaxies. For that, a small set of relations built from an imposed and unique pattern is used. Only baryonic parameters are taken into account. The
cluster of galaxies 1E0657-56, or the so-called Bullet Cluster, is one of the
hottest, most X-ray-luminous clusters known (Tucker, Tananbaum and Remillard, 1998). Chandra
observations by Markevitch et al. (2002) revealed the cluster to be a supersonic
merger. The interpretation that prevails today is that the dissipationless DM accompanying the stellar component has
bypassed the X-ray-emitting hot plasma region during the collision (Fig. 13).
Thus, due to its remarkable morphology, the Bullet Cluster is presumed to be
the best known system in which to test the DM paradigm (Clowe et al.,
2006). More specifically, the repartition of DM has been indirectly
estimated by the combined strong and weak lensing reconstruction method developed by
Brada\u{c} et al. (2005, 2006). Following this view, the dark matter clump revealed
by the strong and weak lensing map is coincident with the collisionless
galaxies but lies ahead of the X-ray-collisional gas. Let us note, however, that gravitational lensing observations can also be explained by two theories that do not require DM: MOG (Browstein and Moffat, 2007) and an extension of MOND (Skordis and Złośnik, 2021).

The issue
also deserves analysis in the framework of the $\kappa{}$ model. As shown in Fig. 13, the
X-ray-emitting hot plasma (in pink), which contains the bulk of the baryonic
Matter, is localised in the central region of the cluster, while most of the
apparent mass as measured by gravitational lensing, i.e., the stellar component plus
DM (in blue), appears on both sides of this bulk.

\begin{center}
\includegraphics[height=260pt, width=210pt]{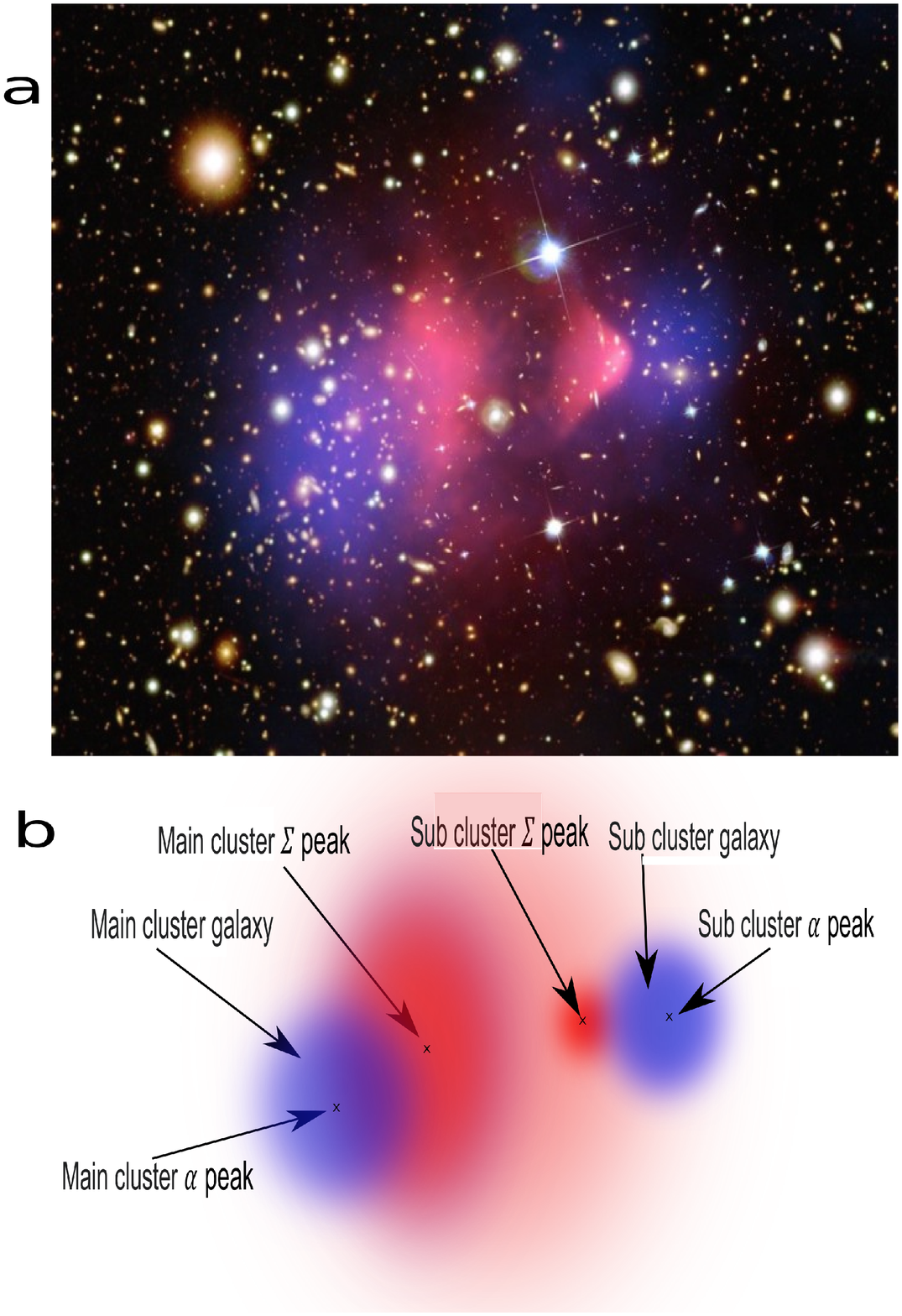}
\end{center}

Fig. 13a. Bullet Cluster (from the Chandra X-ray Observatory: 1E 0657-56), b. the model

\vspace{10pt}
For the Bullet Cluster, as in the preceding study of the individual galaxies
and in accordance with the $\kappa{}$ model, only the intrinsic parameters
linked to the baryonic matter are taken into account.

{\raggedright
The surface density is then modelled by a series of Gaussians as
}

\begin{equation}
\Sigma{}(x,y)= \sum_{i=1}^5 \Sigma{}_i(x,y)=\sum_{i=1}^5{\Sigma{}}_{ci\
}exp\{-\frac{{(x-a_i)}^2}{c_i}-\frac{{(y-b_i)}^2}{d_i}\}
\end{equation}

Each $\Sigma{}_{i}$ is then multiplied by a weighting coefficient
introduced to simulate either the presence of DM or, in
the case of the $\kappa{}$ model, an effect of magnification or even reduction (depending on the location of the observer in a region of either low or high mean density). With this methodology, we fit the profiles of density
presented by Browstein and Moffat (2007) in their full analysis of the Bullet Cluster
(Browstein and Moffat, Fig. 15a and 15b, 2007). The results are displayed in Fig. 14.

{\raggedright

\vspace{3pt} \noindent
\begin{tabular}{p{49pt}p{191pt}p{177pt}}
\parbox{49pt}{\raggedright 
Component
} & \parbox{191pt}{\centering
Main cluster   
} & \parbox{177pt}{\centering
Subcluster
} \\
\hline
\parbox{49pt}{\raggedright

\[
M_g
\]

} & \parbox{191pt}{\raggedright
${\Sigma{}}_{cg1}=0.030, a_{g1}=0.0, b_{g1}=0.0, c_{g1}= 3.3,
d_{g1}=6.6$

${\Sigma{}}_{cg2}=0.045$, $ a_{g2}=0.0, b_{g2}=0.0, c_{g2}=0.25,
d_{g2}= 0.5$
} & \parbox{177pt}{\raggedright

${\Sigma{}}_{cg}=0.035,  a_g=0.7, {b}_g=0.1, c_g=0.05, d_{g1}=0.05$
} \\
\parbox{49pt}{\raggedright

\[
M_{gal}
\]

} & \parbox{191pt}{\raggedright

$
{\Sigma{}}_{cgal}=0.02
, a_{gal}=-0.4, b_{gal}=-0.2, c_{gal}=0.06, d_{gal}= 0.06 $
} & \parbox{177pt}{\raggedright

${\Sigma{}}_{cgal}=0.032, a_{gal}=1.1, b_{gal}=0.2, c_{gal}= 0.06, d_{gal}=0.06$
} \\
\parbox{49pt}{\raggedright

\[
M_{DM}
\]

} & \parbox{191pt}{\raggedright

${p_{cg1}}=2, p_{cg2}=4, p_{cgal}=7$
} & \parbox{177pt}{\raggedright
$p_{cg}=2, p_{cgal}=7$
} \\
\hline
\end{tabular}
\vspace{2pt}

}

{\raggedright
Table 1    DM model:   the surface densities ${
\Sigma{}}_{ci}$ are normalized relative to ${\Sigma{}}_0= 3.1\ {10}^3\ M_{\odot{}}\
{pc}^{-2}$
}

\vspace{10pt}
{\raggedright
In the $\kappa{}$ model, each magnification factor $p_{ci}$ for each
Gaussian component is obtained from the main relationship
}

\begin{equation}
p_{ci}=1+Ln(\frac{{{\Sigma{}}_\odot} {\delta{}}_{ci}}{{\Sigma{}}_{ci}{\delta{}}_\odot})
\end{equation}

where ${\Sigma{}}_{ci}$ and ${\delta{}}_{ci}$ respectively designate the
surface density and the thickness at the maximum of each Gaussian component $i$ and where ${{\Sigma{}}_\odot}=70\ M_{\odot{}}\ {pc}^{-2}$ and ${{\delta{}}_\odot}=500\
pc$ respectively designate the surface density and the mean thickness taken
at the Sun’s position in the Milky Way. These factors are relative to the terrestrial observer; observers located elsewhere would obtain different coefficient values. In particular, Table 2 shows that an estimate by the terrestrial observer gives a value on the order of $6$. Most notably, this is the
same value as for the amplification factor, which is given by Brownstein and Moffat
(2007, Fig. 7) for the gravitational constant. The difference is that in the
$\kappa{}$ model, this factor is calculated, whereas in the MOG framework
proposed by these authors, an extraneous  parametrization is made that modifies $G$. In the $\kappa$ model, the magnification factor results solely from knowledge of the
baryonic density. On the other hand, in the $\kappa{}$-model framework, the gravitational constant $G$, locally
measured by a very distant observer, is the terrestrial Newtonian gravitational constant measured
experimentally on Earth. The relation (14) is valid for the evaluation of the spectroscopic velocities in a galaxy or a galaxy  cluster. This factor corresponds to the usual DM effect as it is perceived from the Earth. However, for gravitational lensing, another set of coefficients must be used; let

\begin{equation}
{p'_{ci}}=1/(1+Ln[\frac{{{\Sigma{}}_{ci}}/{{\delta{}}_{i}}}{\rho_{min}}])
\end{equation}

The reason we follow this procedure is that, in the $\kappa$-model framework, we must necessarily choose a reference point where the distribution of matter is maximized. For a galaxy, this is its centre, which is common to all the distributions for both gas and stars. For the Bullet Cluster, the situation is different, and there are several maximums. Therefore, to determine the lensing diagram, we necessarily use a common reference for the densities. We choose the most natural basis: the baryonic background volume density, $\rho_{min} \sim 10^{-25} \ kg\ m^{-3}$ (Browstein and Moffat, 2007). The factors $p'_{ci}$ are gathered in Table 2.

We must also take into account a global factor at the selected point with coordinates $(x,y)$: $1+Ln[\ \frac{\sum_{i=1}^5\Sigma{}_i(x,y)/{\delta{}_{i}}}{\rho{}_{min}}]$.

The coefficient (15) does not generate a magnification but rather a reduction in the lensing. Indeed, a potential observer living in a very-low-density environment would see an apparent reduction in the lensing and not a magnification. At the same time a significant effect is a displacement of the peaks of the lensing diagram near the centre  of each galaxy cluster as due to the factor $1+Ln[\ \frac{\sum_{i=1}^5\Sigma{}_i(x,y)/{\delta{}_{i}}}{\rho{}_{min}}]$.

{\raggedright

\vspace{3pt} \noindent
\begin{tabular}{p{48pt}p{180pt}p{170pt}}
\parbox{48pt}{\raggedright Component
} & \parbox{180pt}{\centering
Main cluster
} & \parbox{170pt}{\centering
Subcluster
} \\
\hline
\parbox{48pt}{\raggedright

\[
M_g
\]

} & \parbox{180pt}{\raggedright
${\Sigma{}}_{cg1}=0.03, a_{g1}=0.0, b_{g1}=0.0, c_{g1}= 3.3,
d_{g1}= 6.6$

$\delta{}_{g1}=2.6, {p}_{cg1}=8.6, {p'}_{cg}=0.21$

${\Sigma{}}_{cg2}=0.045, a_{g2}=0.0, b_{g2}=0.0, c_{g2}=0.25, d_{g2}=0.5$

${\delta{}}_{g2}=0.7$, $p_{cg2} =6.9$, $p'_{cg2} =0.16$

} & \parbox{180pt}{\raggedright

${\Sigma{}}_{cg}=0.035, a_g=0.7, b_g=0.1, c_g= 0.05, d_g= 0.05$

${\delta{}}_g=0.22, {p}_{cg}=5.9, p'_{cg}=0.14$

} \\
\hline
\parbox{48pt}{\raggedright

\[
M_{gal}
\]

} & \parbox{180pt}{\raggedright

${\Sigma{}}_{cgal}=0.025, a_{gal}=-0.4, b_{gal}=-0.2, c_{gal}= 0.06, d_{gal}=0.06$

${\delta{}_{gal}}=0.24, {p}_{cgal}=6.4, p'_{cgal}=0.14$

} & \parbox{170pt}{\raggedright

${\Sigma{}}_{cgal}=0.032, a_{gal}=1.1, b_{gal}=0.2, c_{gal}= 0.06, d_{gal}=0.06$

${\delta{}}_{gal}=0.24, {p}_{cgal}=6.1, p'_{cgal}=0.14$

} \\
\hline
\end{tabular}
\vspace{5pt}

}

{\raggedright
Table 2       The $\kappa{}$ model: the surface densities ${
 \Sigma{}}_{ci }$ are normalized relative to ${\Sigma{}}_0= 3.1\ {10}^3\ M_{\odot{}}\
{pc}^{-2}$ and the coefficients $\delta{}_i$ (the mean thickness along the line of
sight) to $500 \ kpc$. The coefficient of normalization for the volume density is $\rho_0=3.6\ 10^{-22} \ kg\ m^{-3}$.
}

\vspace{10pt}

The lensing diagram is reported to the terrestrial observer by multiplying by the constant factor $p=1+Ln(\frac{{\Sigma{}}_\odot /{\delta{}}_\odot}{\rho_{min}})$. The latter factor is a simple multiplicative constant and leaves unchanged the orientation of the deviation vectors of light rays and the position of the peaks in the lensing diagram, but it now produces a global magnification $\sim 10$ of the lensing. This set of coefficients must be admitted as empirical factors following a similar protocol as for the individual galaxies. Although there is no theoretical justification at this stage for these empirical formulae, the great interest of the approach is that no extraneous parameters linked to hypothetical DM are introduced in the reasoning.

\begin{center}
\includegraphics[height=260pt, width=210pt]{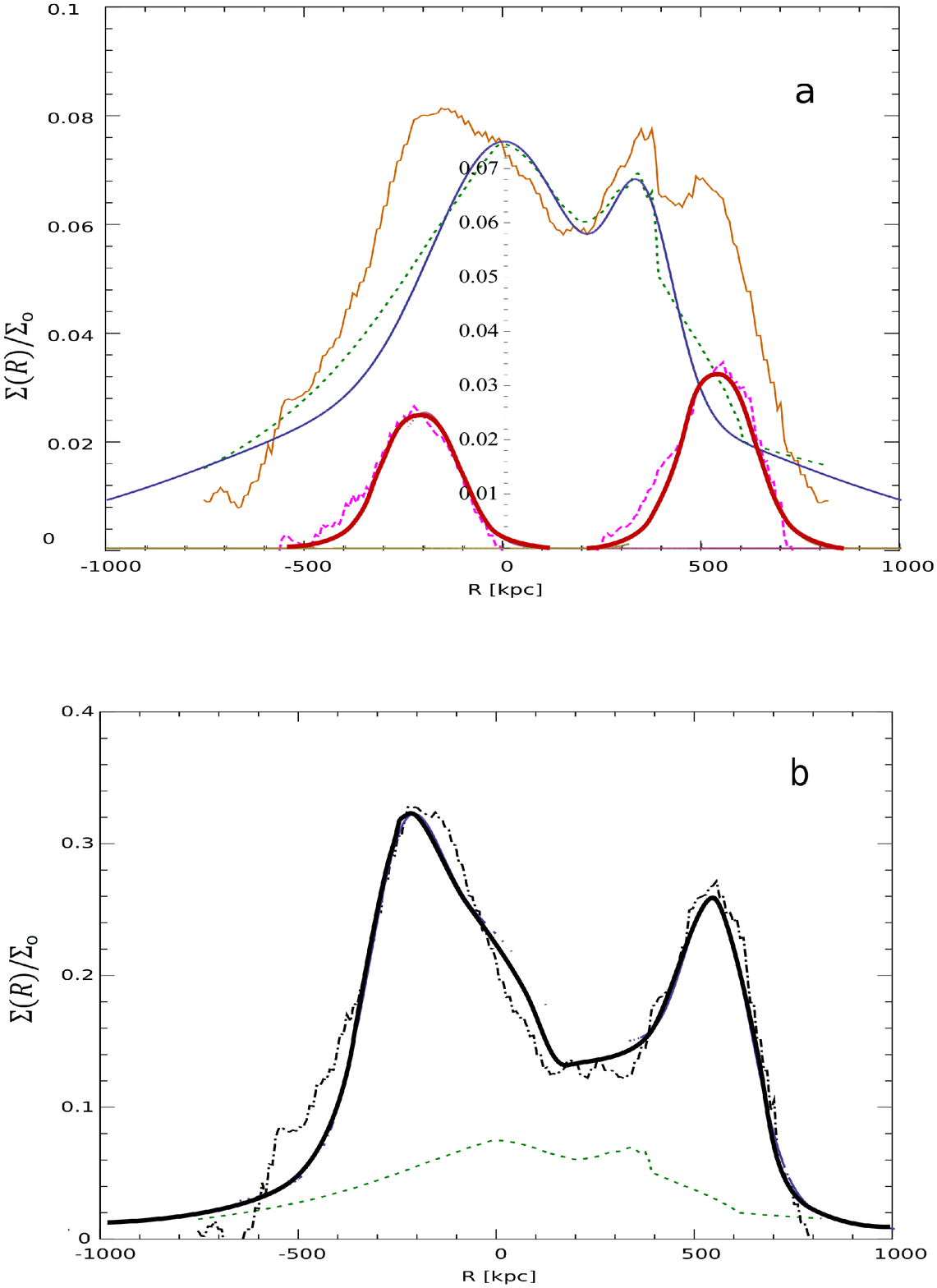}
\end{center}

Fig. 14. Bullet Cluster: a. Plots of the scaled surface densities along the line connecting the galaxy cluster peaks.
The contribution of galaxies is shown in long-dashed magenta, and the total visible baryonic mass is shown in solid brown. The gas distribution is shown in short-dashed green on plots a and b (from Browstein et Moffat, Fig. 15, 2007). Our fits are shown in red for the galaxies and in blue for the gas. b. The DM contribution is shown in dashed-dotted black (from Browstein et Moffat, Fig. 15, 2007). Our fit from Table 1 is in solid black.

\vspace{10pt}
The $\kappa$ model is based on an empirical relationship and is not a theory by itself. A theory must still be chosen as support, for instance, Newtonian dynamics for the rotation curves of individual galaxies and general relativity for the gravitational bending of light by mass. We know that
galaxy clusters can produce noticeable lensing effects. The field
equations of general relativity can be linearised if the gravitational field is
weak. Then, the deflection angle of a set of masses is simply the vectorial sum of
the deflections due to individual lenses. Let $(x,y)$ be the plane of the sky. The deviation angle $\boldsymbol{\alpha{}}$ can be written using the thin lens approximation as (Bartelmann and Schneider, 2001)

\begin{equation}
{\boldsymbol{\alpha{}}}(x,y)=\sum_{i=1}^5 \left[\int_{component\ i}dx'dy'\frac{4G}{c^2}\
\Sigma{}_i\left(x',y'\right)\frac{\mathbf{r}-\mathbf{r}'}{{\left\vert{}\mathbf{r}-\mathbf{r}'\right\vert{}}^2}\right]
\end{equation}

where $G$ is the gravitational constant and $c$ is the speed of light. The
$\kappa{}$ model does not change the local physics, apart from magnification
factors. Thus, in this model, the relation (16) immediately becomes

\begin{equation}
{\boldsymbol{\alpha{}}}(x,y)=p\left\{1+Ln[\ \frac{\sum_{i=1}^5\Sigma{}_i(x,y)/{\delta{}_i}}{\rho{}_{min}}]\right\}\sum_{i=1}^5 {p'_{ci}}
\left[\int_{component\ i}dx'dy'\frac{4G}{c^2}\
\Sigma{}_i\left(x',y'\right)\frac{\mathbf{r}-\mathbf{r}'}{\left\vert{}\mathbf{r}-\mathbf{r}'\right\vert{}^2}\right]
\end{equation}

In Fig. 15, the results for $\left\vert{}{\boldsymbol{\alpha{}}}(x,y)\right\vert{}$ are
displayed for the sole baryonic contribution (Fig. 15a), taking into account 
the DM component (Fig. 15b) and the $\kappa{}$-model
framework (Fig. 15c). Subfigure a shows that in the case of the sole
baryonic contribution, as appropriate, the lensing diagram is centred on the bulk of hot gas
that is the most massive and dense component in the cluster. On the other hand, the lensing is small. This situation
dramatically changes with the introduction of the DM (Fig. 15b) or
within the $\kappa{}$-model framework, where in the latter, the density effect is taken into
account (Fig. 15c). A bipolar configuration then appears with two poles centred
on the galaxies.
Comparison of Fig. 15b and 15c indicates that the lensing diagrams are
similar.

\begin{center}
\includegraphics[height=420pt, width=310pt]{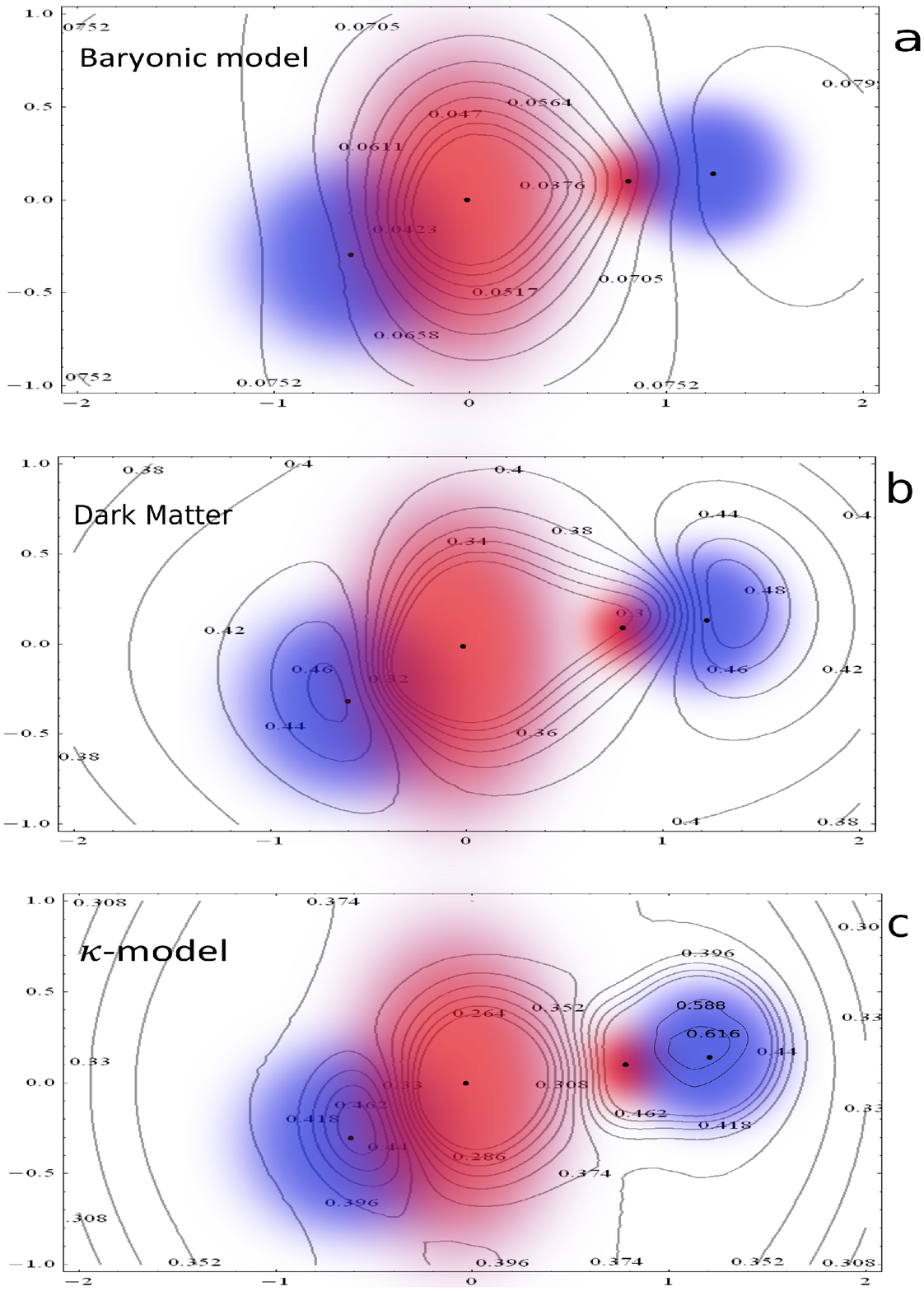}
\end{center}

Fig. 15. Bullet Cluster, lensing diagrams: a. baryons, b. DM, and c. the $\kappa$ model

\section{Conclusion}

We have succeeded in showing that a set of results concerning the dynamics
of individual galaxies or gravitational lensing in clusters of galaxies can be
fitted with one unique component, i.e., baryonic matter. Thus, rather than
invoking exotic particles, a much more efficient path consists of following the
historical order in science: Kepler laws preceded Newtonian mechanics, and the Rydberg formula preceded quantum mechanics. In the present context, coming first are
empirical relationships built on the sole characteristics of baryonic
matter. If this type of relation exists, then a direct link between the baryonic surface densities and the galactic rotation curves is automatically designed. Next, we elaborate a theory supporting and explaining
these relationships following, for instance, MOND or one of its more sophisticated extensions. Similarly, even though the procedure adopted in the present
paper is based upon the use of very simple ``recipes’’, this
strongly suggests that models such as MOND or the $\kappa{}$ model are
privileged against other contingencies that introduce extraneous and
experimentally undetected particles, such as DM or negative masses. A question, however, remains: why can a model that takes into account solely baryonic data be derived at the first level of approximation (smoothing the fluctuations) with rotation curves similar to those of the DM paradigm, which uses extraneous parameters? This question requires further thought. We can assume that the DM effect represents intrinsic properties of the baryonic matter itself. When the acceleration is very low  MOND (Milgrom, 1983) proposed a modification of the inertia term. On the other hand  MOG (Moffat, 2008) predicts an increase in the gravitational constant in the long-range domain. The phenomenological $\kappa$ model (Pascoli, 2022) suggests a renormalization of the dynamics equation with the following consequence: an apparent modification of both inertia and the gravitational constant. Let us also note a promising extension of MOND as developed by Skordis and Złośnik (2021).

On the other hand, the issue is far from being settled at a much higher level of approximation because accurate observational rotation curves
and a precise estimate of the surface densities is then needed to
discriminate between all the competitors: DM, MOG,
MOND or even the $\kappa{}$ model. Unfortunately, at present, the
observational rotation velocity curves very often vary substantially from one
author to another. Another difficulty is the uncertainties in the  mass-to-light ratio. This parameter must be very accurately estimated to deduce a valuable surface mass density; otherwise, the quality of a model cannot be evaluated. With the first objective achieved, the next step in the $\kappa{}$-model framework is to go beyond the first
level of approximation supplied here. The aim is to build a
self-consistent procedure linking the baryonic surface density and the rotation curve
for any galaxy by taking into account the heterogeneities and the variation in both thickness and
inclination, all without any extraneous parameters.

\vspace{10pt}
{\raggedright{\textbf{Data availability statement}: The author confirms that the data supporting the findings of this study are available within the article and the reference list.}}

\vspace{10pt}
{\raggedright{
{\textbf{Conflicts of Interest:} The author declares no conflict of interest.}}}

\section{References}

{\raggedright

Ammazzalorso, S., et al., 2020, Phys. Rev. Lett., 124, 101102

Bartelmann, M., \& Schneider, P., 2001,  Physics Reports, 340,
291

Bertone, G., 2010, Particle Dark Matter: Observations, Models and Searches, Cambridge University Press. p. 762, ISBN 978-0-521-76368-4

Bertone, G., \& Tait, T. M. P., 2018, Nature, 562, 51

Binney, J., \& Merrifield, M., 1998, Galactic Astronomy, Princeton University Press

Boveia A., \& Doglioni, C. , 2018, Annual Review of Nuclear and Particle
Science, 68, 429

Boyarsky, A., Drewes, M., Lasserre, T., Mertens, S., Ruchayskiy, O., 2019, Progress in Particle and Nuclear Physics, 104, 1

Brada\u{c}, M., Schneider, P., Lombardi, M., \&  Erben, T., 2005, A \& A, 436,
39

Brada\v{c}, M.,  Clowe, D.,  Gonzalez, A.H.,  Marshall, P.,  Forman, W.,
 Jones, C.,  Markevitch, M.,  Scott Randall, S., Schrabback, T., \& Zaritsky, D.,
2006, ApJ, 652, 937

Browstein, J.R.,  \& Moffat, J.,  2007, MNRAS, 382,  29

Caputo, R.,  Buckley, M.R., Martin P., Charles, E., Brooks, AM., Drlica-Wagner,
A., Gaskins, J., \& Wood, M., 2016, Phys. Rev. D~\textbf{93}, 062004

Chemin, L.,  Carignan, C.,  \& Foster, T., 2009, ApJ, 705, 1395

Clowe, D.,  Brada\v{c}, M.,  Gonzalez,  A.H., Markevitch, M., Randall,
S.W., Jones, C., \&  Zaritsky,  D., 2006, ApJ,  648, L109

Corbelli, E.,  2003, MNRAS, 342,  199

Corbelli, E., \&  Salucci, P., 2007,  MNRAS, 374, 1051

Famaey, B., \& McGaugh, S.S., 2012, Living Rev. Relativity, 15, 10

Farnes, A., 2018, A \& A., 620, A92

Karukes, E.V.,  \&  P. Salucci, P., 2017, MNRAS, 465, 4703

Lelli, F., Fraternali, F. \& Sancisi, R. 2010, Astronomy \& Astrophysics, 516, A11

L\'{o}pez-Corredoira, M.,  \& Betancort-Rijo J.E., 2021, ApJ, 909,  137

Markevitch, M., Gonzalez, A. H., David, L., Vikhlinin, A., Murray, S., Forman,
W., Jones, C., \& Tucker, W. 2002, ApJ, 567, L27

Mc Gaugh, S., 2015, Canadian Journal of Physics 93, 250

Mc Gaugh, S., 2020, Galaxies, 8, 35

Milgrom, M., 1983, ApJ,  270, 365

Milgrom, M., 2019, arXiv: 1910.04368v3

Milgrom, M., 2020, Studies in the History and Philosophy of Modern Physics, 71, 170

Moffat, J.W., 2006, J. Cosmol. Astropart. Phys., 2006, 4

Moffat, J.W., 2008, Reinventing Gravity, Eds.
HarperCollins

 Pascoli  G., \& Pernas  L., 2020,  hal.archives-ouvertes.fr/hal-02530737

Pascoli, G, 2022, Astrophys. Space Sci., 367, 121

Roszkowski, L.,  Sessolo, E.M., \&   Trojanowski, S., 2018, Rep. Prog. Phys., 81, 066201

 Skordis, C, \&  Złośnik, T., Phys. Rev. Lett., 127, 161302 (2021)

Socas-Navarro, H., 2019, A\&A, 626, A5

Tucker, W.H., Tananbaum, H.,  \&  Remillard, R.A., 1998, ApJ, 496, L5

Xenon Collaboration, 2018, Phys. Rev. Letters 121, 111302.

Zwicky, F.,1933,  Helvetica Physica Acta, 6, 110

}

\end{document}